\documentclass[aps,prl,twocolumn,superscriptaddress,10pt]{revtex4-2}

\usepackage{epsfig}
\usepackage{graphicx}
\usepackage{palatino}
\usepackage[english]{babel}
\usepackage{hyphenat}
\usepackage{amsmath}
\usepackage{amssymb}
\usepackage{mathtools}
\usepackage{mathrsfs}
\usepackage{bbm} 
\usepackage{bm}
\usepackage{slashed}
\usepackage{epstopdf}
\usepackage{xcolor}
\usepackage{booktabs}
\usepackage{float}
\definecolor{lcolor}{rgb}{0.,0.0,0.}
\definecolor{citcolor}{rgb}{0,0.,0.5}
\usepackage[breaklinks,colorlinks,urlcolor=blue,citecolor=blue,linkcolor=blue]{hyperref}
\usepackage{multirow}
\usepackage{ltablex}
\usepackage{soul}

\def\bs{\boldsymbol}

\newcommand{\eqn}[1]{Eq.~\eqref{#1}}

\newcommand{\nn}{\nonumber\\ }

\def\be{\begin{eqnarray*}}
\def\ee{\end{eqnarray*}}
\def\beq{\begin{eqnarray}}
\def\eeq{\end{eqnarray}}

\newcommand{\bea}{\beq \begin{aligned}}
\newcommand{\eea}{\end{aligned}\eeq}

\newcommand{\cO}{{\cal O}}

\newcommand{\Qc}{S}
\newcommand{\Qdip}{S_{\rm dip}}
\newcommand{\rmd}{{\rm d}}

\newcommand{\jet}{{\rm jet}}

\begin{document}

\title{The Colors of Jet Quenching}
\author{Hannah Bossi}
\email{hannah.bossi@cern.ch}
\affiliation{Massachusetts Institute of Technology, Cambridge, MA, 02139, USA}
\author{Maxence Larose}
\email{maxence.larose@stonybrook.edu}
\affiliation{Physics Department, Brookhaven National Laboratory, Upton, NY 11973, USA}
\affiliation{Center for Nuclear Theory, Department of Physics and Astronomy, Stony Brook University, Stony Brook, New York 11794-3800, USA}
\author{Yacine Mehtar-Tani}
\email{mehtartani@bnl.gov}
\affiliation{Physics Department, Brookhaven National Laboratory, Upton, NY 11973, USA}

%\date{\today}

\begin{abstract}

We combine inclusive jet nuclear modification factors with energy--energy correlators to perform a data-driven extraction of the quark and gluon quenching factors and the medium resolution scale from the small-angle region. Taking coherent energy loss as a null hypothesis, we find jet quenching factors incompatible with Casimir scaling. Incorporating color decoherence through antenna energy loss improves the description of the data and restores Casimir scaling, providing evidence that the quark-gluon plasma partially resolves the internal structure of jets in heavy-ion collisions.

\end{abstract}

% insert suggested keywords - APS authors don't need to do this
% \keywords{Perturbative QCD, jet quenching, heavy ion collisions}

%\maketitle must follow title, authors, abstract, and keywords

\maketitle
%%%%%%%%%%%%%%%%%%%%%%%
\section{Data-Driven Approach to Jet Quenching}
%%%%%%%%%%%%%%%%%%%%%%%
Collisions of ultra-relativistic heavy ions provide a unique laboratory for studying the dynamics of Quantum Chromodynamics~(QCD) at high temperatures and energy densities. Over the past two decades, the heavy-ion programs at the Relativistic Heavy Ion Collider~(RHIC) and the Large Hadron Collider~(LHC) have accumulated an unprecedented wealth of data, enabling precision studies of the quark-gluon plasma (QGP) with rarely-produced probes such as QCD jets. The latter are reconstructed from the sprays of energetic particles produced from the fragmentation and hadronization of hard-scattered partons. Measurements of jets have evolved to now encompass both inclusive observables, such as the canonical nuclear modification factor~\cite{ALICE:2023waz,STAR:2020xiv,ATLAS:2018gwx,CMS:2021vui,ATLAS:2022agz,ATLAS:2023iad,ALICE:2024gqr}, and a broad range of jet-substructure observables~\cite{ATLAS:2022vii,ATLAS:2023hso,ATLAS:2025svn,ATLAS:2019dsv,CMS:2025ydi,ALargeIonColliderExperiment:2021mqf,ALICE:2022vsz,ALICE:2024jtb,ALICE:2023dwg,ALICE:2024fip,CMS:2026szl,CMS:2026bmz,CMS:2025dnx,CMS:2024zjn} that reveal modifications to the internal structure of jets. A recent review of jet measurements can be found in Ref.~\cite{Cunqueiro:2021wls}. Complementary measurements of inclusive jets and their substructure are expected to provide precise insight into the underlying QCD dynamics of the QGP, from parton energy loss and medium-induced energy transport to color-coherence effects that are sensitive to the medium resolution scale and, potentially, the quasiparticle structure of the medium. However, to accurately decipher this information from the available data, theoretical frameworks and their application to phenomenology are crucial.  

Substantial progress has been made to advance theoretical calculations \cite{Mehtar-Tani:2010ebp,Mehtar-Tani:2011hma,Mehtar-Tani:2012mfa,Casalderrey-Solana:2011ule,Blaizot:2014bha,Iancu:2000hn,Caucal:2018dla,Caucal:2019uvr,Mehtar-Tani:2017web,Vaidya:2026yfa,Singh:2024vwb,Mehtar-Tani:2024smp,Mehtar-Tani:2025xxd, Caucal:2026lvn,Caucal:2026dsq,Caucal:2021bae,Abreu:2024wka,Arnold:2008zu,Arnold:2020uzm,Feal:2019xfl,Andres:2020vxs} and their phenomenological applications, particularly to jet-substructure observables \cite{Mehtar-Tani:2016aco,Barata:2020rdn,Barata:2021wuf,Barata:2023zqg,Soudi:2025lei,Soudi:2026fls,Pablos:2025cli,Mehtar-Tani:2021fud} beyond the leading-order radiative energy-loss picture \cite{Gyulassy:1990ye,Wang:1991xy,Wang:1992qdg,Kovner:2003zj,Baier:1996kr,Baier:1996sk,Zakharov:1996fv,Zakharov:1997uu} (see also the recent review Ref.~\cite{Mehtar-Tani:2025rty} and references therein). More recently, energy-energy correlators~(EECs) have attracted considerable attention as jet substructure observables due to their reduced sensitivity to soft physics and natural scale separation~\cite{Andres:2022ovj,Andres:2023xwr,Andres:2024ksi,Andres:2024hdd,Barata:2023bhh,Barata:2025fzd,Yang:2023dwc,Singh:2024vwb}.
One commonly utilized approach is to compare experimental data with a variety of parton-shower event generators incorporating different jet--medium interactions \cite{Zapp:2013vla,Zapp:2026cqf,Caucal:2018ofz,Casalderrey-Solana:2014bpa,Chen:2017zte,Schenke:2009vr,Armesto:2009fj,JETSCAPE:2017eso}. Such an approach has proven to be very useful for gaining a qualitative understanding of the data, but has limitations when providing a quantitative description that isolates a specific jet--medium interaction. This is primarily due to the fact that models include several difficult-to-control approximations applied to multiple different physics components. 

So-called ``selection bias'', which we define as an apparent jet-substructure modification resulting from gluons losing more energy than quarks, also makes it more difficult to isolate effects of genuine substructure modifications~\cite{CMS:2024zjn,Qiu:2019sfj,Du:2020pmp}. Such modification may be attributed to the medium resolving the subjet color charges down to an angular resolution scale~$\theta_{\rm c}$ \cite{Mehtar-Tani:2011hma,Casalderrey-Solana:2012evi}, an effect commonly referred to as color-decoherence. Below that scale, the medium is ``blind'' to the subjet inner structure \cite{Mehtar-Tani:2010ebp,Mehtar-Tani:2011hma,Mehtar-Tani:2012mfa,Casalderrey-Solana:2011ule,Mehtar-Tani:2017ypq}. Recent phenomenological models compared to experimental data show indications that support the existence of such a transition~\cite{ATLAS:2022vii,ATLAS:2025svn,Kudinoor:2025gao}. Other phenomenological approaches instead aim to describe energy loss solely in terms of modifications to the quark and gluon fractions~\cite{Spousta:2015fca,Pradhan:2025tla,CMS:2020plq,Qiu:2019sfj}, and yield a reasonable description of experimental data. This Letter presents a data-driven investigation of the color-decoherence transition and the quark- and gluon-jet quenching factors, combining theoretical input and information extracted from different jet observables in the small-angle region, where theoretical control is expected to be better. 

For precision phenomenology, analytic calculations tailored to specific observables within well-controlled regions of phase space provide an effective strategy to probe medium properties: they yield robust and systematically improvable predictions, and they avoid global analyses that drive fitted parameters into regimes where the underlying approximations no longer hold. They can also guide the development of the next generation of general-purpose Monte Carlo event generators.

The central thesis of this Letter is that the existing body of measurements can be used to test fundamental hypotheses, such as coherent jet energy loss, and to extract physically meaningful quantities, such as the quark- and gluon-jet quenching factors. 

We emphasize the need to combine inclusive and substructure observables such as the inclusive jet production cross-section and the EECs. As such, this work aims to provide a potentially more robust and clearer path toward identifying the QCD dynamics governing jet--medium interactions.

The data-driven methodology we propose begins by testing the null hypothesis of coherent jet energy loss before introducing color-decoherence effects based on a formulation reported in a companion paper~\cite{Mehtar-Tani:2026}. This hierarchical approach allows individual hypotheses to be assessed with minimal reliance on modeling and ensures that additional complexity is introduced only when required by the data. It thereby provides a more robust basis for discriminating among physical mechanisms and for guiding future phenomenological developments.

%%%%%%%%%%%%%%%%%%%%%%%%%%%%%%%%%%%%%%%%%%%%%%%%%%%%%%%%%%%%%%%%%%%%%
\section{The null hypothesis: coherent energy loss}
%%%%%%%%%%%%%%%%%%%%%%%%%%%%%%%%%%%%%%%%%%%%%%%%%%%%%%%%%%%%%%%%%%%%%
We hypothesize that at sufficiently small jet opening angle $R$, the medium does not resolve the jet substructure, leaving it unmodified apart from a shift in the parent jet transverse momentum ($p_{\rm T}$) due to the energy loss of its total color charge. A similar hypothesis was adopted in Ref.~\cite{Qiu:2019sfj} to extract flavor-dependent jet-energy-loss distributions from inclusive jet nuclear modification factors. However, such an extraction faces an identifiability problem: this class of inclusive observables cannot independently constrain quark- and gluon-jet modifications without additional assumptions. The resulting uncertainties may therefore partly reflect the rigidity of the model rather than the constraining power of the data.

A genuinely data-driven determination of the flavor dependence of jet quenching therefore requires at least one additional observable with a different sensitivity to quark- and gluon-initiated jets. More generally, one would consider at least two observables, $\cO^{(1)}$ and $\cO^{(2)}$, receiving distinct quark and gluon contributions:
\begin{align}\label{eq:syst-equations}
\cO^{(1)}
&=
\Qc_q\, \cO_q^{(1)}
+
\Qc_g\, \cO_g^{(1)}\,,
\nn
\cO^{(2)}
&=
\Qc_q\, \cO_q^{(2)}
+
\Qc_g\, \cO_g^{(2)}\,.
\end{align}
The two measurements provide independent information only if their relative sensitivities to quark and gluon jets differ.

We first apply this framework to proton--proton ($pp$) collisions using the measured inclusive jet cross section~\cite{ATLAS:2018gwx} and EEC distribution~\cite{CMS:2025ydi}, with the same kinematic selections as the heavy-ion measurement considered below.

{\it  Proton--proton baseline fits}: The inclusive jet cross section in $pp$ collisions for jets reconstructed with cone size $R$ can be decomposed into quark- and gluon-initiated contributions, $\rmd \sigma_\jet = \rmd \sigma_q+\rmd \sigma_g$.
This decomposition naturally defines the quark and gluon jet fractions at a given $p_{\rm T}$, $f_q
=\rmd \sigma_q/ \rmd \sigma_\jet$ and $f_g
=\rmd \sigma_g/ \rmd \sigma_\jet$, 
which satisfy $f_q+f_g=1$. 

The EEC measures the angle between two particles inside the jet and sums over all pairs and may be normalized by the jet yield \cite{Basham:1978zq,Basham:1978bw},  
\beq
\frac{\rmd \Sigma}{\rmd R_{\rm L}  \rmd p_{\rm T}} 
\simeq \left(
\frac{\rmd \sigma_{\rm jet}}{\rmd p_{\rm T}}
\right)^{-1}
\underset{i\neq j}{\sum}
\int \rmd \sigma
\frac{p_{\rm T,i} p_{\rm T,j}}{p_{\rm T}^2}
\delta\left(
R_{\rm L}-\theta_{ij}
\right),
\eeq
where $\rmd \sigma$ denotes the differential cross section, including the squared matrix element and the corresponding phase-space measure. The EEC admits a similar decomposition into quark and gluon  contributions, 
$\rmd \Sigma_\jet = f_q\, \rmd \Sigma_q+  f_g \, \rmd \Sigma_g\,$. 

In this work, we employ the leading-logarithmic resummation formula (LL) for the EEC derived in \cite{Dixon:2019uzg}. The same result is independently re-derived using generating-functional methods in a companion paper \cite{Mehtar-Tani:2026}. In addition, we parametrize the transition to the non-perturbative region by replacing the running-coupling scale with a flavor-dependent infrared-regulated scale,
$\mu^2\rightarrow \left(\mu^{\beta}+m_{q,g}^{\beta}\right)^{2/\beta}\,$.
The infrared scales $m_{q,g}$ and the interpolation parameter $\beta$ are determined from the $pp$ data. Also, we multiply the EEC distribution by an overall $K$-factor, which parametrizes corrections to the normalization not captured at LL accuracy. Taking the quark and gluon fractions from \textsc{Pythia} simulations, the remaining four free parameters are determined through a simultaneous fit to the CMS EEC data~\cite{CMS:2025ydi}, 
yielding $K = 2.51$, $\beta=5.53$, $m_q=2.16$~GeV and $m_g=4.21$~GeV (see Supplemental Material for additional details). This simple model can be systematically refined beyond the scope of the present proof-of-concept analysis, for instance by incorporating higher-order perturbative corrections and a more detailed treatment of the non-perturbative transition.

{ \it Joint fit of $R_{\rm AA}$ and EEC:} 
Turning to heavy-ion collisions, we account for quenching in both the inclusive jet cross section and the EEC. We first consider the nuclear modification factor and define the flavor-dependent quenching factors as \cite{Mehtar-Tani:2024mvl,Mehtar-Tani:2025rty}
\beq
&&S_{q,g}(p_{\rm T})   = \left(
\frac{\rmd \sigma_{\rm pp}^{q,g}}{\rmd p_{\rm T}}
\right)^{-1}
\int_0^\infty \rmd\epsilon\, 
P_{q,g}(\epsilon,p_{\rm T})\,
\frac{\rmd \sigma_{\rm pp}^{q,g}}{\rmd p_{\rm T}'}\,,
\eeq
where $p_{\rm T}'=p_{\rm T}+\epsilon$ and $P_{q,g}(\epsilon, p_{\rm T})$ denotes the quark or gluon energy-loss distributions \footnote{At leading-logarithmic accuracy, this distribution can be related to the jet function evaluated at the natural jet scale $\mu=p_{\rm{T}} R$, namely $S_i\sim J_i(\mu=p_{\rm T} R)$ \cite{Mehtar-Tani:2024mvl,Mehtar-Tani:2025rty}.}.

If valid, the coherent-energy-loss hypothesis would allow us to directly constrain the flavor dependence of jet quenching through the quark- and gluon-jet nuclear modification factors, $R_{\rm AA}^{q}\equiv \Qc_q$ and $R_{\rm AA}^{g}\equiv \Qc_g$, by applying \eqn{eq:syst-equations} with $\cO^{(1)}\equiv \rmd\sigma_{\rm jet}$ and $\cO^{(2)}\equiv \rmd\Sigma_{\rm jet}(R_{\rm L})$.

The EEC data exhibit three distinct features: an enhancement at small angles, followed by a suppression below unity and a subsequent rise at large angles. The latter is commonly attributed to soft medium-induced radiation and medium response \cite{Andres:2022ovj,Barata:2023bhh,Yang:2023dwc} and is not included in our framework. Motivated by the LPM suppression of medium-induced radiation at angles below the coherence angle, we hypothesize that, around and below this scale, the soft-mode dynamics responsible for the large-angle rise is subleading. The region below the dip should therefore be dominated by hard modes whose color charges lose energy either coherently or independently, depending on whether they are resolved by the medium. In the absence of soft particles, the Pb--Pb-to-$pp$ EEC ratio is also expected to decrease with increasing angle in Monte Carlo simulations~\cite{Barata:2023bhh,Bossi:2024qho}. To minimize sensitivity to soft dynamics, we restrict our fits to small angles, typically below the minimum of the ratio at $R_{\rm L}\sim 0.1$, and vary this upper cutoff to assess the associated systematic uncertainty.

The quark and gluon contributions determined from the $pp$ baseline are reweighted by their respective quenching factors, $\Qc_q$ and $\Qc_g$, and simultaneously fitted to the ATLAS $R_{\rm AA}$~\cite{ATLAS:2018gwx} and CMS EEC data~\cite{CMS:2025ydi} for $R=0.4$ jets, yielding a reasonable description of both observables.
One could stop the analysis at this stage. However, the quality of the fit alone does not validate the coherent-energy-loss hypothesis, which must be subjected to further consistency tests. A more significant tension emerges from the extracted flavor dependence: the independently determined quark- and gluon-jet quenching factors, $\Qc_q\simeq 0.8$ and $\Qc_g\simeq 0.3$, are incompatible with the expected Casimir (color charge) scaling (see Supplemental Material for details).
Neither geometric averaging nor higher-order shower effects are expected to account for a discrepancy of this magnitude, as both generally reduce the contrast between quark- and gluon-jet quenching \cite{Apolinario:2020nyw}.  

This illustrates an important point: agreement with the measured distributions alone does not necessarily validate the underlying physical picture. By combining observables with complementary sensitivity to quark and gluon jets, the fit can expose inconsistencies in the inferred microscopic parameters that would remain hidden in either observable separately. In the present case, the anomalously strong quark--gluon hierarchy provides a nontrivial test of the coherent-energy-loss hypothesis and motivates the inclusion of corrections associated with the medium resolving the jet substructure.

Recent theoretical developments have identified systematic corrections to the coherent limit, arising at next-to-leading order when a collinear splitting forms a two-parton antenna that loses energy as a color-correlated system~\cite{Mehtar-Tani:2017ypq,Mehtar-Tani:2011lic}. These effects have been resummed at LL accuracy, successfully describing jet $R_{\rm{AA}}$~\cite{Mehtar-Tani:2017web,Mehtar-Tani:2021fud} and azimuthal-anisotropy data~\cite{Mehtar-Tani:2024jtd,Pablos:2025cli}. More recently, effective-field-theory approaches have provided a systematic treatment of these interference effects to all orders~\cite{Mehtar-Tani:2024smp,Mehtar-Tani:2025xxd,Vaidya:2026yfa,Caucal:2026dsq,Caucal:2026lvn}.

%%%%%%%%%%%%%%%%%%%%%%%%%%%%%%%%%%%%%%%%%%%%%%
\section{Color-decoherence corrections and Casimir scaling}
%%%%%%%%%%%%%%%%%%%%%%%%%%%%%%%%%%%%%%%%%%%%%%
Before turning to a quantitative analysis, let us first discuss the qualitative features expected from the gradual onset of color decoherence, largely independently of the details of any particular model or theoretical framework. The transition between the coherent and decoherent regimes is governed by a characteristic angular scale $\theta_{\rm c}$. In the multiple-soft-scattering approximation for a static medium, this scale behaves parametrically as $\theta_{\rm c}\sim(\hat q L^3)^{-1/2}$. Collinear splittings at angles below $\theta_{\rm c}$ remain unresolved by the medium, and the resulting partonic system loses energy coherently as a single color charge corresponding to its parent parton. At larger angles, the medium resolves the individual color charges, leading to enhanced energy loss.

This picture has two immediate consequences. First, decoherence increases the suppression of both quark- and gluon-initiated jets relative to the fully coherent limit. 
Second, because narrow configurations lose less energy than wider, resolved configurations, the selected jet sample becomes biased toward smaller opening angles. The EEC is consequently enhanced at small angles and depleted at larger angles. 
This angular-selection effect compounds the modification caused by selection bias. 

Before discussing the medium-modified EEC let us introduce quenching factors of partonic pairs. For instance,  $\Qc_{qg}(\theta)$ denotes the quenching factor of a hard collinear quark--gluon pair with opening angle $\theta$. In the large-$N_c$ limit,  it can be approximated as
$\Qc_{qg}(\theta)\simeq \Qc_q\,\Qdip$,  
where $\Qdip(\theta)$ describes the additional suppression associated with the resolved color dipole \cite{Mehtar-Tani:2017ypq}. We neglect the angular dependence of the primary-quark quenching factor, which is justified for collinear splittings with $\theta\ll R$. Recall that quenching factors are related to Laplace transforms of energy loss distributions~\cite{Caucal:2026lvn}. 

The dipole quenching factor $\Qdip(\theta)$  is derived in Ref.~\cite{Mehtar-Tani:2017ypq}~(see Supplemental Material for the explicit expression). 
It approaches unity in the coherent limit, $\theta\ll\theta_{\rm c}$, because the medium does not resolve the two color charges. In the opposite limit, $\theta\gg\theta_{\rm c}$, the quark and antiquark lose energy independently, and the dipole quenching factor approaches the product of their individual quenching factors, $\Qdip(\theta)\to \Qc_q \Qc_{\bar q}\simeq \Qc_q^2$. As we shall see, the functional form of the interpolation between these two limits plays a crucial role in describing the data and provides a useful lever arm for discriminating among different descriptions of color decoherence. In particular, a sharp transition at $\theta_{\rm c}$ can be contrasted with the smooth QCD-based result of Ref.~\cite{Mehtar-Tani:2017ypq}, whose residual coherent component exhibits a long power-law tail proportional to $\theta^{-2/3}$, so color-coherence effects persist to angles above $\theta_{\rm c}$. In the multiple-soft-scattering approximation the medium resolution angle reads 
\beq\label{eq:res-angle-c}
\theta_{\rm{c}}=\sqrt{\frac{12}{\hat q L^3}}\,,
\eeq
where $\hat q$ is the jet quenching parameter \cite{Mehtar-Tani:2025rty}. In this analysis, we neglect higher-order corrections to the quenching factors and assume $S_q(\theta>\theta_{\rm c})\simeq S_q(R)$ \cite{Caucal:2026dsq}. To first approximation, this allows us to identify the extracted jet quenching factors $S_{q,g}$ with the corresponding subjet quenching factors entering the dipole quenching factor. 

Consequently, even for a relatively small value of $\theta_{\rm c}$, residual color-coherence effects persist to angles parametrically larger than the nominal coherence angle.

Following Ref.~\cite{Mehtar-Tani:2026}, a jet-calculus approach based on generating functionals was used to derive the medium-modified EEC. This framework naturally incorporates both color coherence and the decoherent limit at large $N_c$, assuming that the quenching factors depend only weakly on the subjet energies. The key observation is that the contribution associated with the total color charge can be factorized, leaving only dipole quenching factors in the perturbative expansion of the EEC. The approach applies to hard-collinear modes with energies well above the characteristic medium scales and is therefore expected to remain controlled only at sufficiently small angles. At larger angles, medium-induced radiation and medium response are expected to become important. Our goal in this work is to restrict the analysis to a theoretically controlled angular region in which these effects are subleading, thereby improving the accuracy with which medium properties—such as the flavor-dependent quenching factors and the medium resolution angle—can be extracted.

The evolution equation of the EEC in the medium remains the same as in vacuum except for a modification of the anomalous dimension  \cite{Mehtar-Tani:2026}
\begin{align}\label{eq:eec-ll_med}
{\bs \Sigma}^{\rm med}
&=(1,1) \cdot \exp\left[ -\frac{1}{4\pi} \int_{R_{\rm L}}^R  \frac{\rmd \theta^2}{\theta^2 }\alpha_s(p_{\rm T}\theta ) \, {\bs \gamma}_{\rm med}(\theta)\right]\,,
\end{align}
for the cumulative distribution where ${\bs \Sigma}\equiv (\Sigma_q,\Sigma_g)$. The EEC distribution is obtained by taking the derivative of \eqn{eq:eec-ll_med} with respect to $R_{\rm L}$. The medium anomalous dimension reads
\begin{align}
{\bs \gamma}_{\rm med}(\theta) =
\begin{pmatrix}
\frac{25}{6}C_F \, \Qdip(\theta) & -\frac{7}{15}n_f \\[4pt]
-\frac{7}{6}C_F\,\Qdip (\theta)& \frac{14}{5}C_A \,\Qdip(\theta)+ \frac{2}{3}n_f
\end{pmatrix}\,.
\end{align}

The $n_f$ terms remain unmodified: the contribution to $\gamma_{gg}$ is purely virtual and already included in the factored-out gluon quenching weight, while $\gamma_{qg}$ is $1/N_c$ suppressed and neglected at large $N_c$. For the EEC, each flavor contribution is 
weighted by both the total gluon and quark charge quenching factors and the corresponding 
dipole quenching factor. Schematically, 

\beq
\Sigma_{\rm jet}=
\frac{f_q\,\Qc_q\,\,\Sigma^{\rm med}_q(R_{\rm L})
+
f_g\,\Qc_g\,\,\Sigma^{\rm med}_g(R_{\rm L})}{f_q \Qc_q + f_g \Qc_g}\,.\nn
\eeq

\noindent For the $R_{\rm AA}$ we simply have
\beq \label{eq:raa}
R_{\rm AA} = f_q \Qc_q + f_g \Qc_g\,.
\eeq

\begin{figure}
    \begin{center}
    \includegraphics[width=0.4\textwidth]{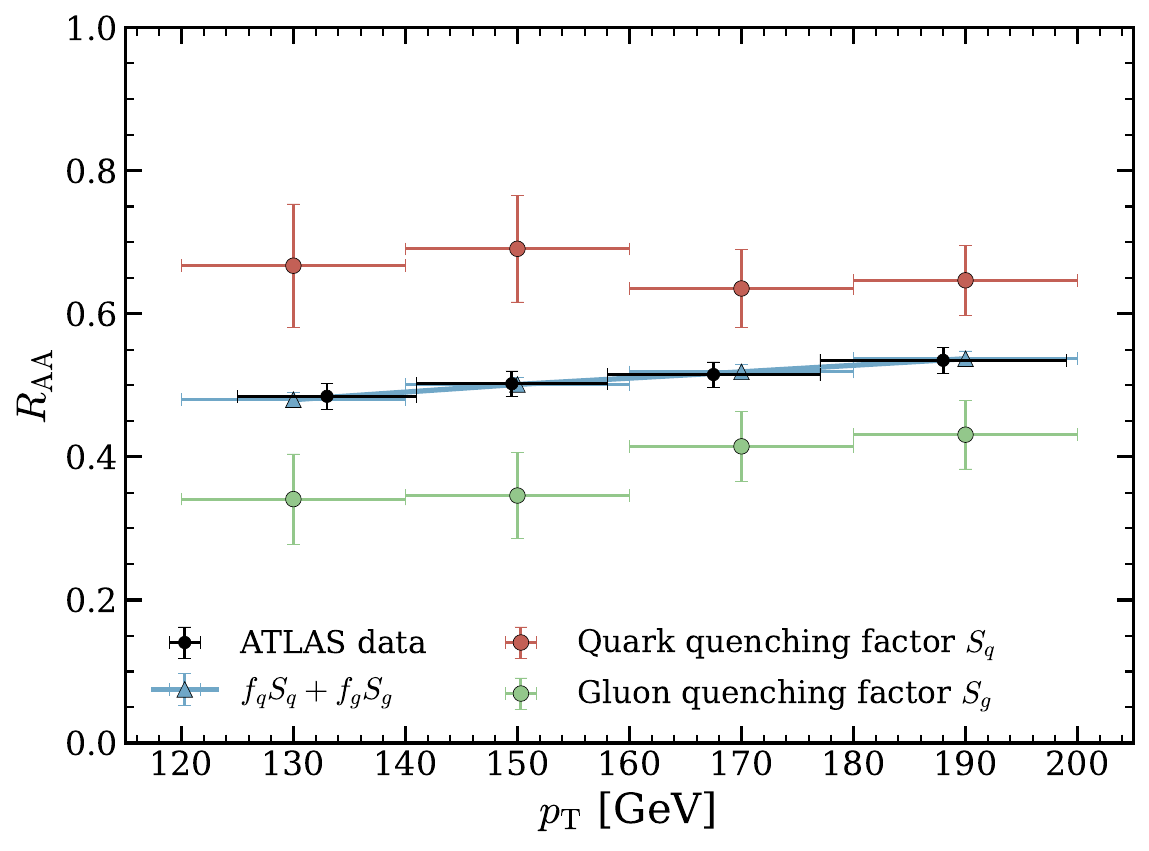} 
    \end{center}
    \caption{Fit to the ATLAS jet nuclear modification factor, together with the quark- and gluon-jet quenching factors,~$S_q$ and~$S_g$. The
    quenching factors are extracted from a joint fit to the ATLAS $R_{\rm{AA}}$~\cite{ATLAS:2018gwx} and the CMS EEC~\cite{CMS:2025ydi}, taking color-decoherence corrections into account. In contrast to the purely coherent approximation, the extracted factors are compatible with Casimir scaling, $S_g \simeq S_q^{9/4}$ (see Supplemental Material). Horizontal error bars span the $p_{\rm T}$ window. The vertical error bars on the ATLAS points are their statistical and systematic uncertainties added in quadrature; those on $S_q$ and $S_g$ are the corresponding fit uncertainties.
    }
    \label{fig:atlas_raa_decoh}
\end{figure}

\begin{figure}
    \begin{center}
    \includegraphics[width=0.4\textwidth]{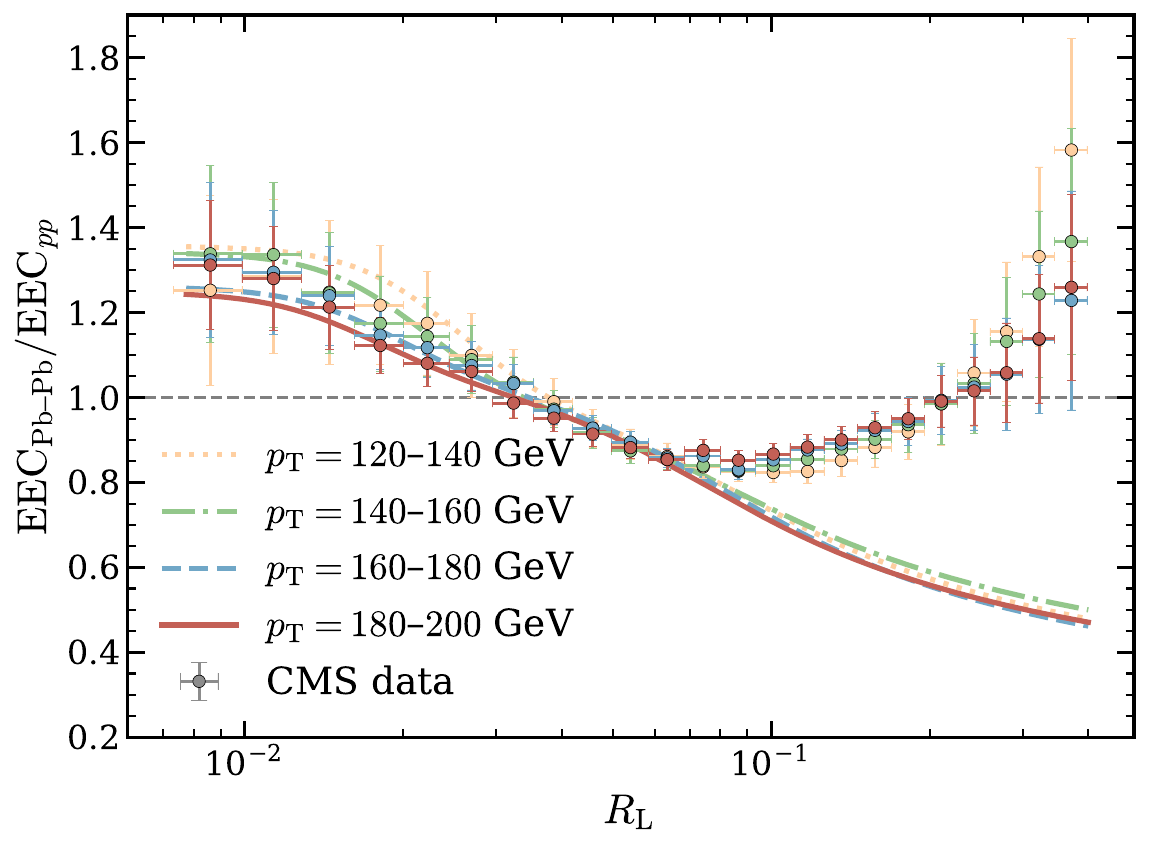} 
    \end{center}
    \caption{CMS measurement of the Pb--Pb to $pp$ EEC ratio~(points)~\cite{CMS:2025ydi}, compared with the fit that takes color-decoherence corrections into account~(curves). Only points with $R_{\rm L}<0.07$ are fitted, restricting the analysis to the small-angle region where hard-collinear dynamics are expected to dominate. The systematic uncertainty associated with this cutoff is assessed in the Supplemental Material by varying $R_{\rm L}^{\max}$ between $0.05$ and $0.10$.} 
    \label{fig:cms_eec_pbpb_pp_decoh}
\end{figure}

We perform again the joint fit of  $R_{\rm AA}$ and the EEC and the results are shown in Figs.~\ref{fig:atlas_raa_decoh} and \ref{fig:cms_eec_pbpb_pp_decoh}. Overall, this approach provides a better description of the combined data ($\chi^2/\mathrm{dof}\simeq 0.13$) compared to the coherent case ($\chi^2/\mathrm{dof}\simeq 0.93$). The absolute $\chi^2/\mathrm{dof}$ values do not measure the precision of the calculation, since the published uncertainties are added in quadrature, with their correlations unavailable to us. The comparison instead shows the relative quality of two fits sharing the same vacuum reference, flavor fractions and angular window. Moreover, the fits yield two remarkable results that support the robustness of our theoretical framework. First, the extracted quark and gluon quenching factors, $\Qc_q\simeq 0.65$ and $\Qc_g\simeq 0.40$, are compatible with Casimir scaling, which predicts $\Qc_q^{9/4} \simeq 0.38$~(see Supplemental Material). These values are compatible with other previous extractions for quark and gluon quenching factors~\cite{Spousta:2015fca}. 

\begin{figure}
    \begin{center}
    \includegraphics[width=0.4\textwidth]{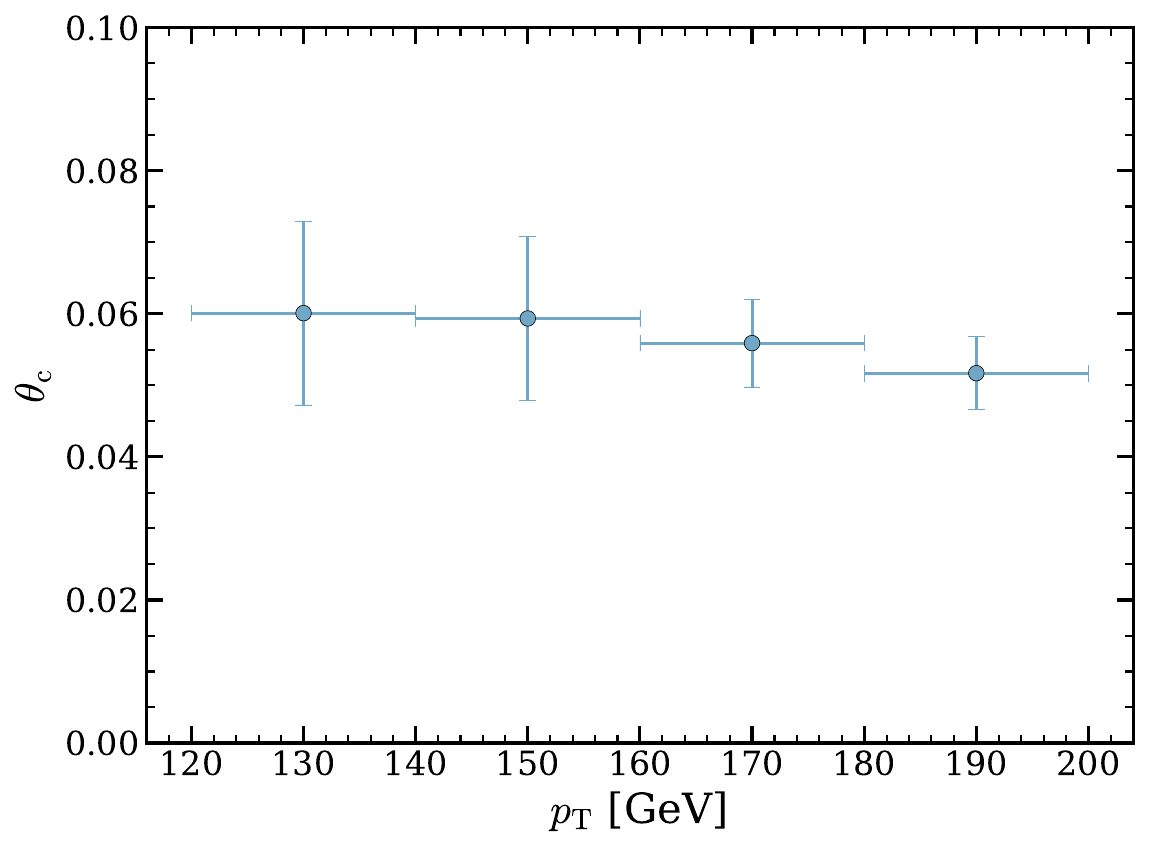} 
    \end{center}
    \caption{Fitted decoherence angle $\theta_{\rm c}$ as a function of $p_{\rm T}$. As expected, no significant $p_{\rm T}$ dependence is observed, and the extracted values are consistent with a characteristic angle~$\theta_{\rm c} \simeq 0.05-0.07$. The vertical error bars are only related to the uncertainty associated with the fitting procedure. 
    }
    \label{fig:thetaC}
\end{figure}

Second, the medium resolution angle is remarkably well constrained with an average value over all four $p_{\rm T}$ windows of
\beq 
\theta_{\rm c} = 0.057 \pm 0.005~({\rm fit})\,^{+0.014}_{-0.008}~({\rm syst})\,,
\eeq
as shown in Fig.~\ref{fig:thetaC}. The quoted uncertainties on that figure are only those associated with the fitting procedure. Details about the systematic uncertainties are provided in the Supplemental Material. Furthermore, relating the extracted $\theta_{\rm c}$ to a quenching parameter requires modeling the collision geometry and space-time evolution of the plasma. We leave both aspects for future work.

It is worth noting that the quark-only distribution in Fig.~\ref{fig:atlas_raa_decoh} shows similar quenching to that observed in the $\gamma$-tagged $R_{\rm AA}$ from ATLAS~\cite{ATLAS:2023iad}, which is expected to be quark-dominated. This agreement provides an important independent consistency check of the framework and of the extracted quark-quenching factor. More generally, a global analysis incorporating a broader set of jet-substructure observables would provide a more stringent test of this picture, while consistently accounting for correlations among the extracted parameters. We leave such an analysis for future work.

\section{Toward the extraction of medium properties}
\label{sec:conclusion}

In this Letter, we propose a data-driven approach for both testing QCD-based descriptions of jet quenching and extracting quantities such as the color-coherence angle $\theta_{\rm c}$ and the quark and gluon quenching factors, $S_q$ and $S_g$. Our methodology restricts the extraction to a theoretically well-controlled region of phase space, namely the small-angle region, where we hypothesize that color coherence dominates and medium-induced gluon radiation, suppressed by the LPM effect, and medium-response effects are subleading. Here, we do not aim for a global extraction; rather, this work serves as a first proof of concept for the approach.

Within the fully coherent picture, both observables can be described, but the extracted quark and gluon quenching factors are incompatible with Casimir scaling. Including color decoherence resolves this tension while simultaneously determining a characteristic medium resolution angle. More generally, our results demonstrate that combining observables with complementary flavor sensitivities can test the physical consistency of jet-quenching mechanisms, rather than merely their ability to reproduce individual measurements.

Several extensions provide natural directions for future work. The centrality dependence of $\theta_{\rm c}$~\cite{Mehtar-Tani:2024jtd} could provide further information on the medium resolution scale, while forthcoming sPHENIX data and available EEC measurements from STAR~\cite{STAR:2025jut} will enable analogous studies at RHIC energies, where the quark-initiated jet fraction is significantly larger. Extending the calculation to higher perturbative accuracy and incorporating a broader set of observables, such as the groomed jet radius, would pave the way toward a more systematic global analysis. Ultimately, combining such theory-constrained analyses with modern machine-learning tools and Monte Carlo event generators may provide a path toward precision extractions of QGP properties.

\vspace{0.5cm}
\begin{acknowledgments}
\noindent{\bf Acknowledgments.} The authors would like to thank Paul Caucal, Chris McGinn, and Peter Steinberg for insightful discussions related to this work. Y.~M.~T. was supported by the U.S. Department of Energy under Contract No. DE-SC0012704. H.B. was supported by the U.S. Department of Energy under grant Contract
Number DE-SC0011088.
\end{acknowledgments}

%\appendix
%%%%%%%%%%%%%%%%%%%%%%%%%%%%%%%%%%%%%%%%%%%%%%
\section*{Supplemental Material}

\makeatletter
\renewcommand{\p@subsection}{}
\makeatother

%%%%%%%%%%%%%%%%%%%%%%%%%%%%%%%%%%%%%%%%%%%%%%
\subsection{The singlet antenna (dipole) quenching factor}
\label{app:dipole}
%%%%%%%%%%%%%%%%%%%%%%%%%%%%%%%%%%%%%%%%%%%%%%
The energy loss distribution for a singlet antenna was derived in \cite{Mehtar-Tani:2017ypq} based on the assumption of independent multiple medium-induced emissions. 
Making the dependence on the medium length $L$ explicit, the single-charge quenching factor is exponential, $\Qc_q(L)=\exp(-\Gamma_{\rm rad} L)$, with $\Gamma_{\rm rad}$ the radiation rate. The dipole quenching factor can then be written as
\beq
&& \Qc_{\rm dip}(L,\theta)
=
\Qc_q^2(L)
\nn
&&+
2\Gamma_{\rm rad}
\int_0^L \rmd t
\Qc_q^2(L-t)
\exp\left[
-\left(\frac{\theta}{\theta_{\rm{c}}}\right)^2
\left(\frac{t}{L}\right)^3
\right].
\label{eq:dipole-quenching}
\eeq
Here,
\beq\label{eq:res-angle}
\theta_{\rm{c}}^2=\frac{12}{\hat q L^3},
\eeq
in the multiple-soft-scattering approximation where $\hat{q}$ is the jet quenching parameter, and the exponential factor represents the survival probability for color coherence up to time $t$.

To express the result entirely in terms of $\Qc_q$, we use $\Gamma_{\rm rad} L=-\ln \Qc_q$ and introduce the dimensionless time variable $\tau=t/L$. Equation~\eqref{eq:dipole-quenching} then becomes
\beq
&& \Qdip(\theta)
=
\Qc_q^2\nn
&& \times\left\{1- 
2\ln \Qc_q
\int_0^1 \rmd\tau 
\exp\left[
- 2\ln \Qc_q \tau
- \left(\frac{\theta}{\theta_{\rm{c}}}\right)^2\tau^3
\right]\right\}.\nn
\label{eq:dipole-quenching-dimensionless}
\eeq
This expression explicitly reproduces the coherent limit,
\beq
\Qdip(\theta)\to 1
\qquad\text{as}\qquad
\theta\to 0.
\eeq
More importantly, at large angles the dipole quenching factor approaches the fully decoherent limit through a power law rather than exponentially. Although the coherence-survival probability is exponentially suppressed at fixed time, the integration over early splitting times generates a much slower algebraic decay: 
\beq
\Qdip(\theta)- \Qc_q^2
\simeq
-\frac{2\Gamma(1/3)}{3} \, 
\Qc_q^2\ln \Qc_q
\left(\frac{\theta_{\rm{c}}}{\theta}\right)^{2/3}\,,
\eeq
for $\theta\gg\theta_{\rm{c}}$.
Consequently, even for a relatively small value of $\theta_{\rm{c}}$, residual color-coherence effects persist to angles parametrically larger than the nominal coherence angle.
%%%%%%%%%%%%%%%%%%%%%%%%%%%%%%%%%%%%%%%%%%%%%%
\subsection{The EEC to leading logarithmic accuracy} \label{app:eec_resum}
%%%%%%%%%%%%%%%%%%%%%%%%%%%%%%%%%%%%%%%%%%%%%%

The factorization formula for the EEC reads \cite{Dixon:2019uzg}
\begin{align}
&\frac{\rmd \Sigma_{q,g}}
     {\rmd R_{\rm L}}
=
\int_0^1 \rmd x \,
x^2 \,
\frac{\rmd J_{q,g}(x,R_{\rm L},\mu)}
     {\rmd R_{\rm L}}
\,H_{q,g}(x,\mu)\,.
\end{align}
We work at leading-logarithmic~(LL) accuracy, for which the hard and the jet functions $H$ and $J$ can be evaluated at $x=1$. The factorization scale is chosen  $\mu= \, p_{\rm T} \, R$, such that the DGLAP evolution between the natural jet scale $R_0\sim 1$ and the jet radius $R$ is entirely absorbed into the hard function. Consequently, $H_{q(g)}$ can be identified with the quark (gluon)-initiated jet cross section,

\begin{align}
H_{q(g)}(x,\mu=p_{\rm T} R)
\simeq
\delta(1-x)\frac{\rmd \sigma_{q(g)}}{\rmd p_{\rm{T}}}\,.
\end{align}

\noindent At LL accuracy the jet function reads (for $\mu=R p_{\rm{T}}$):
\begin{align}\label{eq:eec-ll}
\frac{\rmd \bs J}{\rmd R_{\rm L}}
=
\frac{\alpha_s(R_{\rm L} p_{\rm{T}})}{2\pi R_{\rm L}}
\,(1,1)\cdot  \left[
\left(
\frac{\alpha_s(R p_{\rm{T}})}
     {\alpha_s(R_{\rm L} p_{\rm{T}})}
\right)^{\frac{{\bs \gamma}(3)}{\beta_0}}
\cdot
{\bs \gamma}(3) \right]\, ,
\end{align}
where ${\bs J}\equiv ({J}_q,{J}_g)$ and
\begin{align}
{\bs \gamma}(3) =
\begin{pmatrix}
\frac{25}{6}C_F & -\frac{7}{15}n_f \\[4pt]
-\frac{7}{6}C_F & \frac{14}{5}C_A + \frac{2}{3}n_f
\end{pmatrix},
\end{align}
denotes the third Mellin moment of the leading-order timelike DGLAP anomalous-dimension matrix, governing the coupled evolution of the quark and gluon jet functions, and $\beta_0= (11 C_A - 2n_f)/3$. In the present work, we normalize the EEC by the inclusive cross section, which at LL accuracy is proportional to $H_q+H_g$, and therefore adopt the notation $J\to\Sigma$.

%%%%%%%%%%%%%%%%%%%%%%%%%%%%%%
\subsection{Two-step fitting procedure}
\label{app:fit-res}
%%%%%%%%%%%%%%%%%%%%%%%%%%%%%%

The fit proceeds in two steps: first, the vacuum reference is fixed against the $pp$ data alone, and then, the medium parameters are extracted with that reference held fixed. The two steps use different upper angular limits, for reasons that are described below.

\begin{figure}[H]
    \begin{center}
    \includegraphics[width=0.4\textwidth]{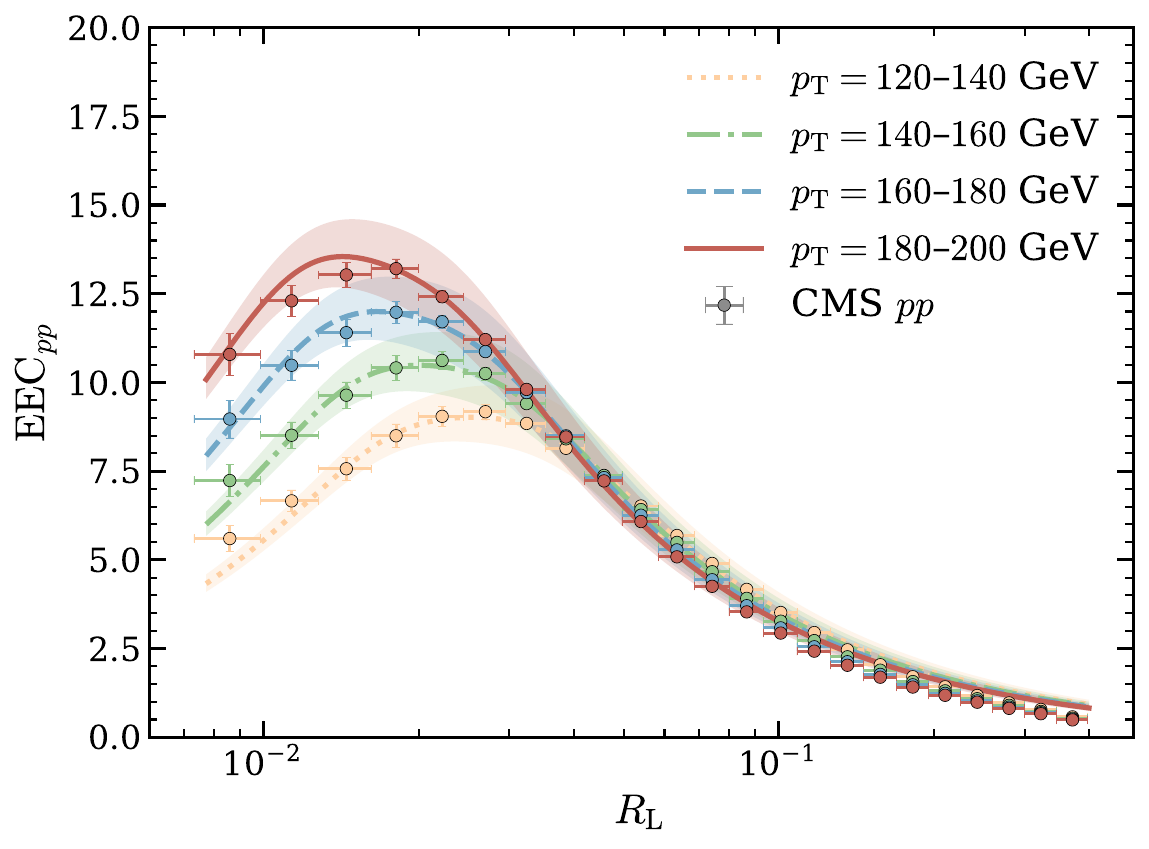} 
    \end{center}
    \caption{Fit to the CMS $pp$ EEC~\cite{CMS:2025ydi} in four jet transverse-momentum intervals spanning $120 < p_{\rm T} < 200$~GeV. The fit is bounded by $R_{\rm{L}} < 0.10$ as the LL description is expected to apply to $R_{\rm{L}} \ll R$ where jet-boundary effects and higher-order matching can be neglected. The fitted parameters are $K = 2.514 \pm 0.008$,
    $\beta = 5.53 \pm 0.36$, $m_q = 2.16 \pm 0.03$~GeV, $m_g = 4.21 \pm 0.07$~GeV. These values are held fixed in every medium fit that follows. The bands represent renormalization-scale variations by a factor of two about the central scale, which entails varying the jet radius from $R/2$ to $2 R$.}
    \label{fig:cms_eec_pp}
\end{figure}

\begin{figure}
    \centering
    \includegraphics[width=0.4\textwidth]{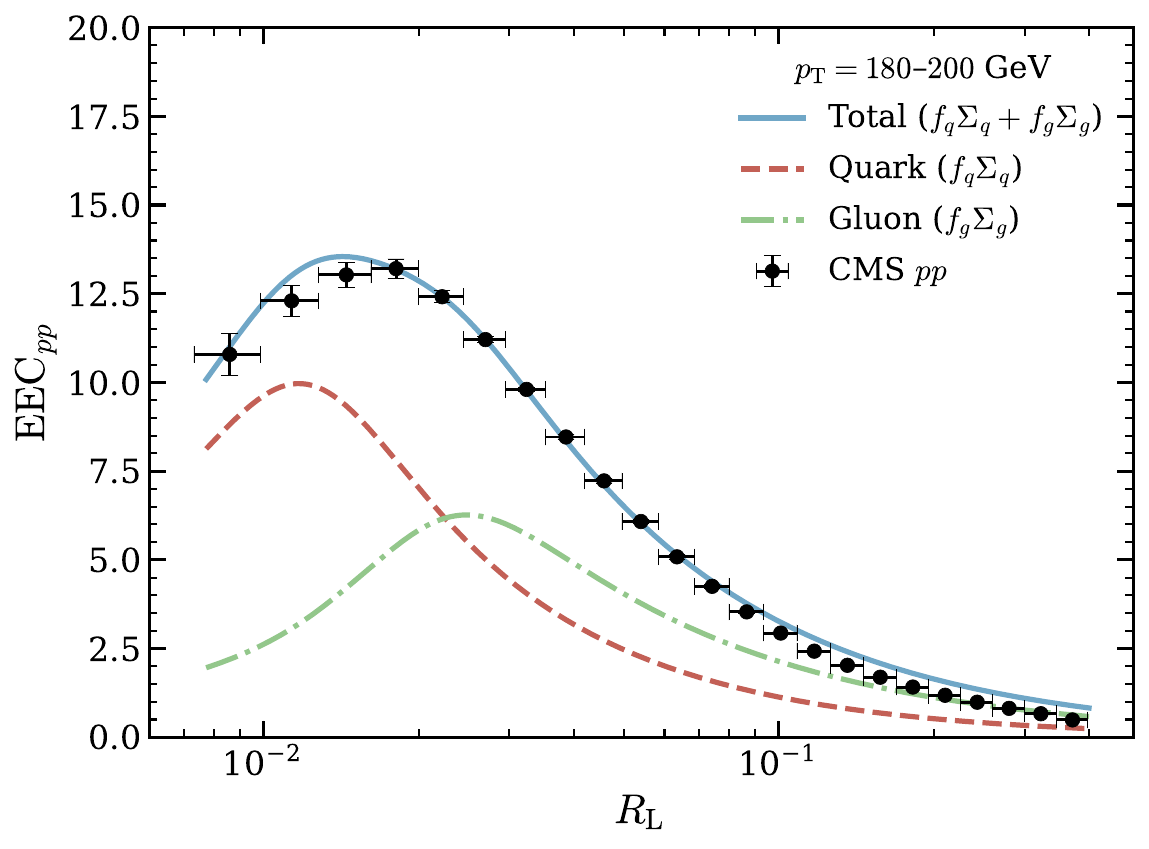}
    \caption{Flavor decomposition of the vacuum EEC in the $180 < p_{\rm T} < 200$~GeV window, compared with the CMS $pp$ measurement~\cite{CMS:2025ydi}. The two lower curves are the quark and gluon correlators weighted by their fractions, $f_q \Sigma_q$ and $f_g \Sigma_g$ with $f_q = 0.492$ and $f_g = 0.508$; the upper curve is their sum, which is what the measurement constrains. Gluon jets radiate more and their correlator peaks at a larger angle, $R_{\rm{L}} = 0.025$ against $0.012$ for quark jets. It is this difference in shape that lets the EEC constrain the quark--gluon composition of the sample, which the inclusive cross section alone cannot.}
    \label{fig:eec_pp_flavor}
\end{figure}

{\it Vacuum step:} The normalization $K$ and the three parameters of the infrared-regulated coupling scale, $\beta$, $m_q$ and $m_g$, are fitted to the CMS $pp$ EEC~\cite{CMS:2025ydi} over the four $p_{\rm T}$ windows ($120<p_{\rm T}<140~\mathrm{GeV}$, $140<p_{\rm T}<160~\mathrm{GeV}$, $160<p_{\rm T}<180~\mathrm{GeV}$, and $180<p_{\rm T}<200~\mathrm{GeV}$), using the points with $R_{\rm L} < 0.10$ and the published uncertainties added in quadrature. The upper angular limit is imposed by the accuracy of the calculation: the LL resummation holds for $R_{\rm L} \ll R$, where jet-boundary effects and higher-order matching may be neglected. Fitting out to the last published point pushes $\chi^2/$dof above $100$, and the
distorted reference then inverts the flavor hierarchy the medium step is meant to extract. The fit returns $K = 2.514 \pm 0.008$, $\beta = 5.53 \pm 0.36$, $m_q = 2.16 \pm 0.03~{\rm GeV}$, and $m_g = 4.21 \pm 0.07~{\rm GeV}$ with $\chi^2/$dof $= 2.09$ for $48$ degrees of freedom, and describes all four windows at once, as shown in Fig.~\ref{fig:cms_eec_pp}. These four numbers are held fixed thereafter, so every medium fit shares one vacuum baseline. Note that quark and gluon jets differ in the shape of their EEC and not only in its normalization: the gluon EEC peaks at
$R_{\rm L} = 0.025$, against $0.012$ for the quark one (Fig.~\ref{fig:eec_pp_flavor}). This is what lets the EEC constrain the flavor composition of the sample, which the inclusive cross section alone cannot.

\begin{figure}[H]
    \centering
    \includegraphics[width=0.4\textwidth]{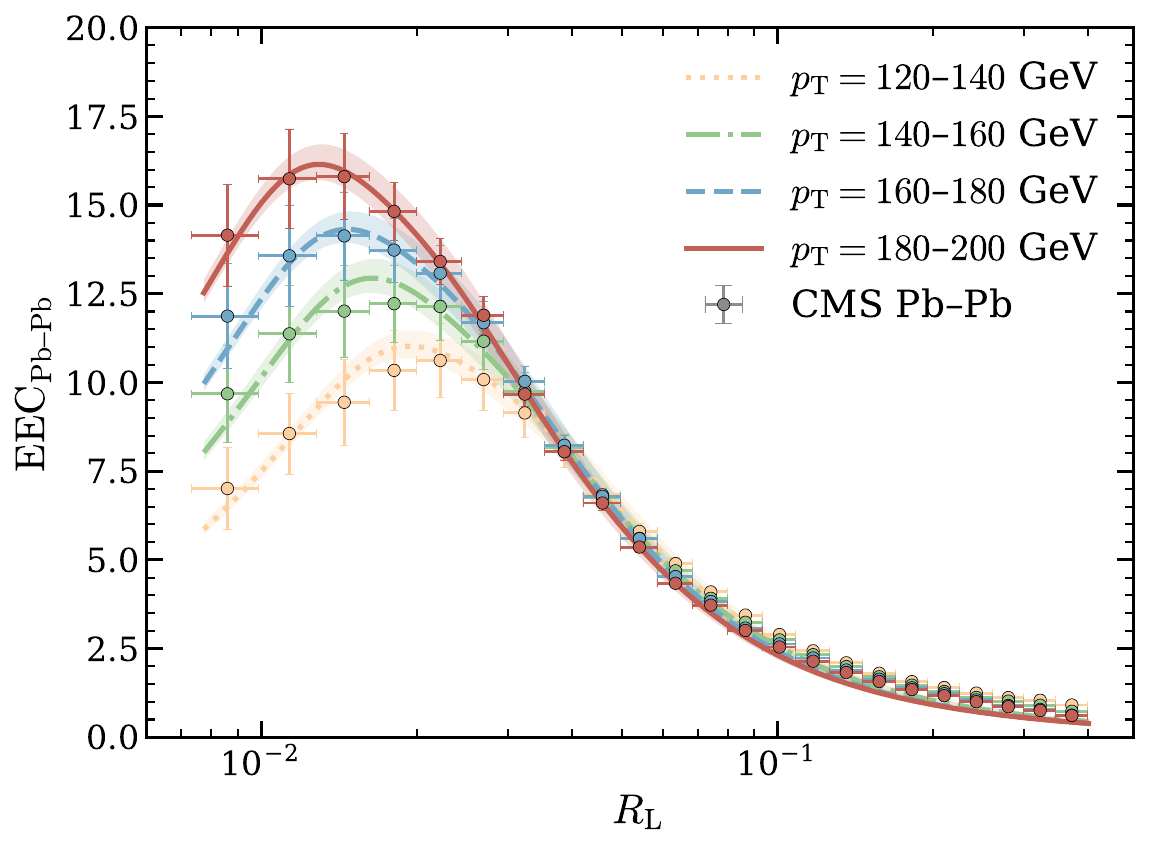}
    \caption{Fit to the CMS Pb--Pb EEC~\cite{CMS:2025ydi} in $0$--$10\%$ central collisions, taking color-decoherence corrections into account. The EEC and the ATLAS $R_{\rm{AA}}$~\cite{ATLAS:2018gwx} are
    fitted jointly, bounded by $R_{\rm{L}} < 0.07$ as in Fig.~\ref{fig:cms_eec_pbpb_pp_decoh}, giving
    $\chi^2/$dof $= 0.13$. Bands are defined as in Fig.~\ref{fig:cms_eec_pp}.}
    \label{fig:eec_pbpb_decoherent}
\end{figure}

\begin{figure}[H]
    \begin{center}
    \includegraphics[width=0.4\textwidth]{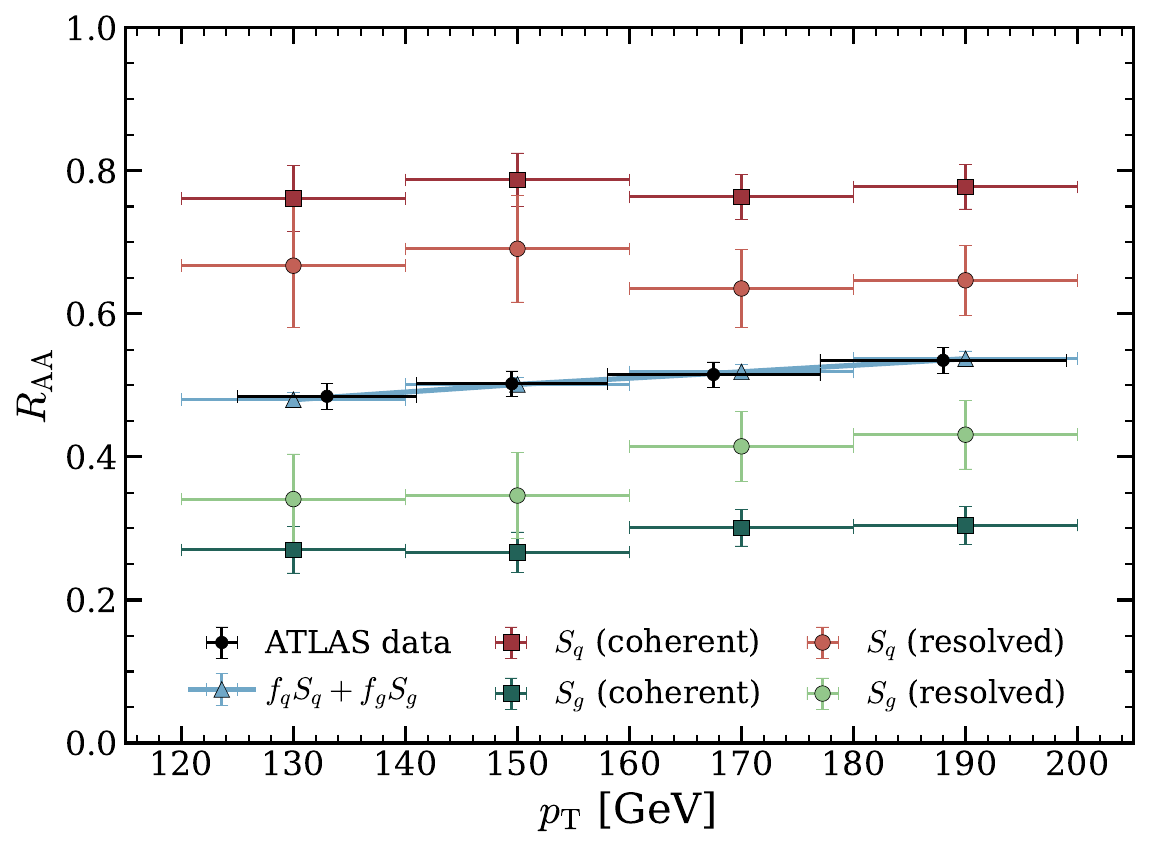} 
    \end{center}
    \caption{Comparison of the quenching factors $S_q$ and $S_g$ extracted in the coherent limit with those obtained by taking color-decoherence corrections into account~(labeled as ``resolved'' in the figure). Both come from a simultaneous fit to the ATLAS $R_{\rm{AA}}$~\cite{ATLAS:2018gwx} and the CMS EEC~\cite{CMS:2025ydi}, and both reproduce the measurement through the same flavor-fraction weighted sum, but they split it very
    differently between the flavors: the coherent fit gives $S_g = 0.27$--$0.30$ against the Casimir
    expectation $S_q^{9/4} = 0.54$--$0.58$, a discrepancy of $8$--$11\sigma$, whereas the resolved fit agrees with it within $1.5\sigma$. Uncertainties as in Fig.~\ref{fig:atlas_raa_decoh}.}
    \label{fig:atlas_raa}
\end{figure}

\begin{figure}[H]
    \centering
    \includegraphics[width=0.4\textwidth]{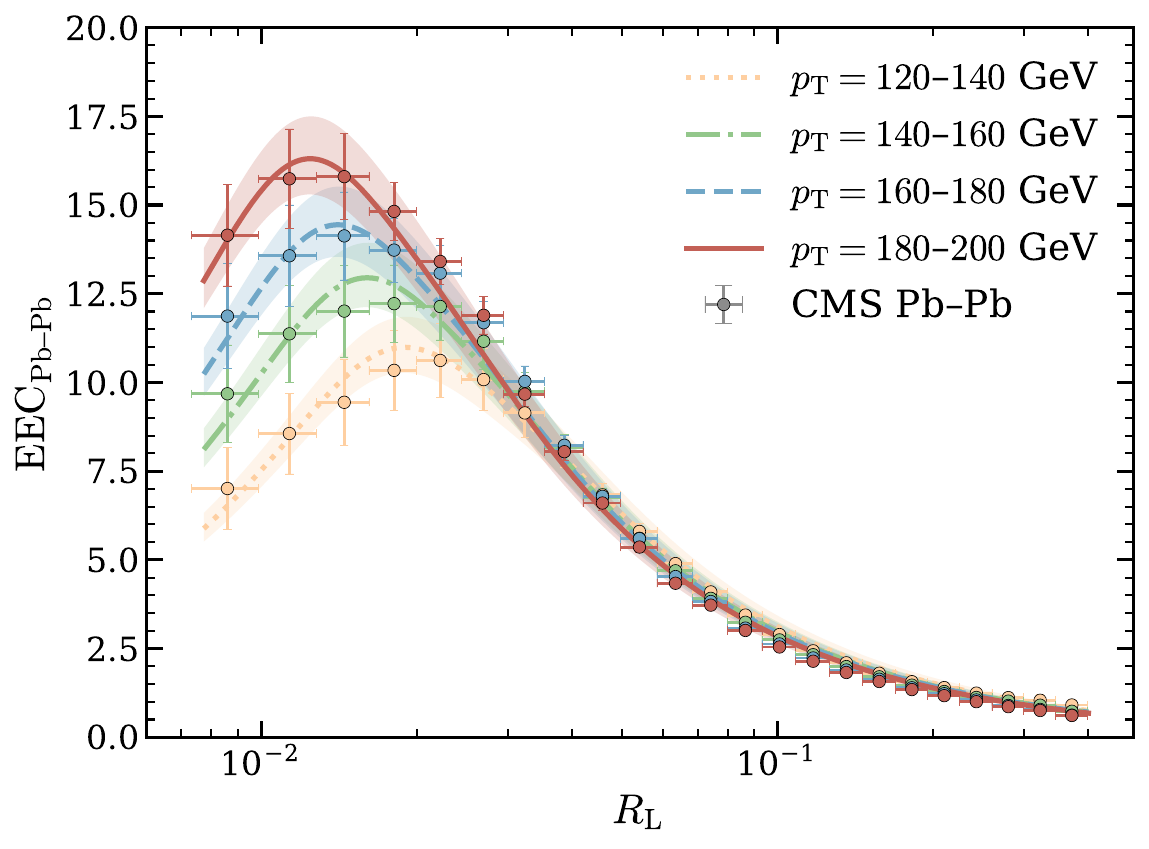}
    \caption{Fit to the CMS Pb--Pb EEC~\cite{CMS:2025ydi} in $0$--$10\%$ central collisions, in the coherent limit. The
    EEC and the ATLAS $R_{\rm{AA}}$~\cite{ATLAS:2018gwx} are fitted jointly, bounded by $R_{\rm{L}} < 0.07$
    as in Fig.~\ref{fig:cms_eec_pbpb_pp}, giving $\chi^2/$dof $= 0.93$. Bands are defined as in
    Fig.~\ref{fig:cms_eec_pp}.}
    \label{fig:eec_pbpb_coherent}
\end{figure}

\begin{figure}[H]
    \begin{center}
    \includegraphics[width=0.4\textwidth]{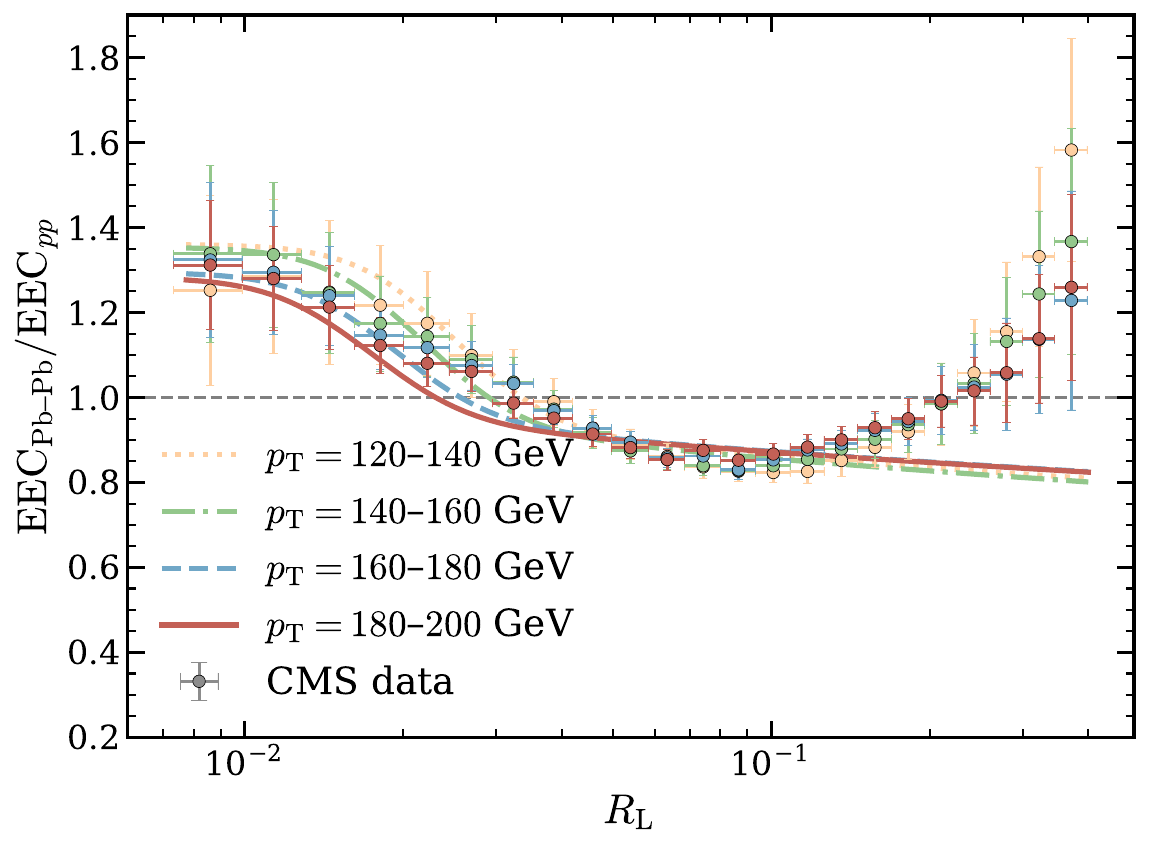} 
    \end{center}
    \caption{CMS measurement of the Pb--Pb to $pp$ EEC ratio~(points)~\cite{CMS:2025ydi}, compared with the fit in the coherent limit~(curves). Only the points with $R_{\rm{L}} < 0.07$ are fitted, as in Fig.~\ref{fig:cms_eec_pbpb_pp_decoh}. The fit gives $\chi^2/$dof $= 0.93$, against $0.13$ by taking color-decoherence corrections into account.
    }
    \label{fig:cms_eec_pbpb_pp}
\end{figure}

{\it Medium step:} The quenching factors $\Qc_q$ and $\Qc_g$, together with $\theta_{\rm c}$ when
color-decoherence corrections are included, are fitted independently in each $p_{\rm T}$ window, to the CMS Pb--Pb EEC~\cite{CMS:2025ydi} and the ATLAS $R_{\rm AA}$~\cite{ATLAS:2018gwx} jointly, using the points with $R_{\rm L} < 0.07$. This limit is a property of the medium model rather than of the vacuum one: it sits just above the fitted decoherence angle, beyond which the medium resolves the dipole and the soft physics the calculation leaves out takes over. Its cost is quantified in App.~\ref{app:rlmax}. The coherent limit gives
$\chi^2/$dof $= 0.93$, with $\Qc_q = 0.76$--$0.79$ and $\Qc_g = 0.27$--$0.30$ across the four windows
(Figs.~\ref{fig:eec_pbpb_coherent} and~\ref{fig:cms_eec_pbpb_pp}); including color decoherence gives $0.13$,
with $\Qc_q = 0.64$--$0.69$, $\Qc_g = 0.34$--$0.43$ and $\theta_{\rm c} = 0.052$--$0.060$
(Fig.~\ref{fig:eec_pbpb_decoherent}), the values window by window being those of
Figs.~\ref{fig:atlas_raa_decoh} and~\ref{fig:thetaC}. Both fits reproduce the measured $R_{\rm AA}$ equally
well, since both are free to choose the sum $f_q \Qc_q + f_g \Qc_g$, but they divide it between the flavors very differently, as shown in Fig.~\ref{fig:atlas_raa}. Indeed, Casimir scaling relates the two quenching factors by $\Qc_g = \Qc_q^{9/4}$, as derived in App.~\ref{app:Casimir}. The coherent limit is not compatible with it: it returns $\Qc_g = 0.27$--$0.30$ where its own $\Qc_q$ implies $0.54$--$0.58$, a discrepancy of $8$--$11\sigma$. Including color decoherence removes the tension without imposing anything, the free fit landing within $1.5\sigma$ of the Casimir expectation in every window; refitting under the constraint rather than testing it afterwards leads to the same conclusion (App.~\ref{app:casimirfit}).

Two caveats remain. First, the quoted uncertainties are those of the medium step alone, at fixed vacuum reference and fixed flavor fractions; a complete treatment would propagate those as well. Second, the published statistical and systematic uncertainties are added in quadrature, their correlations not being available to us, so the
absolute $\chi^2$ values do not measure the precision of the calculation. What they do measure is the relative quality of two fits performed in exactly the same way with the same vacuum reference, flavor fractions and angular window.

\subsection{Monte Carlo determination of the quark and gluon jet fractions}

The quark and gluon vacuum fractions $f_q$ and $f_g$ are obtained from \textsc{Pythia}~8~\cite{Bierlich:2022pfr} simulations of $pp$ collisions at $\sqrt{s} = 5.02$~TeV. All leading-order $2 \to 2$ QCD scattering processes are enabled, and the resulting final-state particles are clustered into jets with the anti-$k_{\rm T}$ algorithm. The simulation is validated against the ATLAS inclusive jet cross section~\cite{ATLAS:2018gwx} and the CMS EEC data~\cite{CMS:2025ydi}. It adopts their acceptance and binning: full jets with $R = 0.4$ and $|y| < 2.8$, binned in four windows spanning $120 < p_{\rm T} < 200$~GeV. The EEC is built from charged tracks with $p_{\rm{T}} > 1$~GeV.

The flavor of each jet is determined by \emph{descendancy tracing}, a tagging algorithm we developed for this
analysis, which proceeds as follows. The shower of each hard-scattering parton is walked forward, generation by generation. A descendant may vote for the flavor of its parton if it lies within $\Delta R < R$ of the jet axis and is a constituent of the jet, or an ancestor of one. Each vote is weighted by the $p_{\rm T}$ of the particle casting it, and the jet is assigned the flavor whose votes have the largest average $p_{\rm T}$. Because only forward daughter relations are used, color reconnection cannot corrupt the assignment.

\begin{figure}[H]
    \centering
    \includegraphics[width=0.4\textwidth]{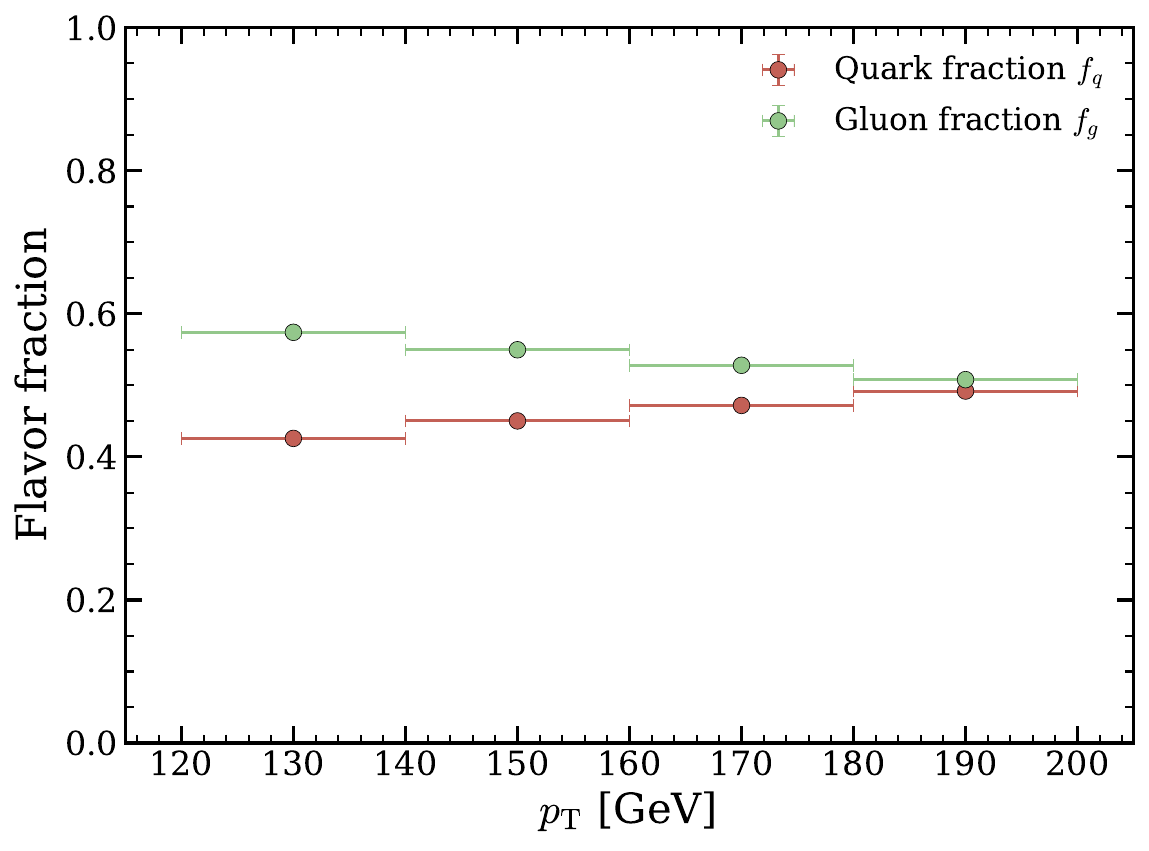}
    \caption{Quark and gluon fractions of the inclusive $pp$ jet cross section obtained from \textsc{Pythia}~8~\cite{Bierlich:2022pfr} simulations for full jets with $R = 0.4$. The flavor of each jet is determined by \emph{descendancy tracing}, a tagging algorithm we developed for this analysis. Horizontal bars span the $p_{\rm T}$ window; the vertical bars give the scatter across the five independent runs of each window, but they are smaller than the markers and therefore not visible.}
    \label{fig:flavor_fractions}
\end{figure}

Every window is generated as five independent runs of $10^6$ events. The fractions quoted in Fig.~\ref{fig:flavor_fractions} are the average over these runs, and their scatter determines the statistical uncertainty of the Monte Carlo event generator. However, this uncertainty is negligible, amounting to at most $0.2\%$ of the fractions in any window.

\subsection{Sensitivity to the decoherence angle}
\label{app:thetacdep}

\begin{figure}[H]
    \centering
    \includegraphics[width=0.4\textwidth]{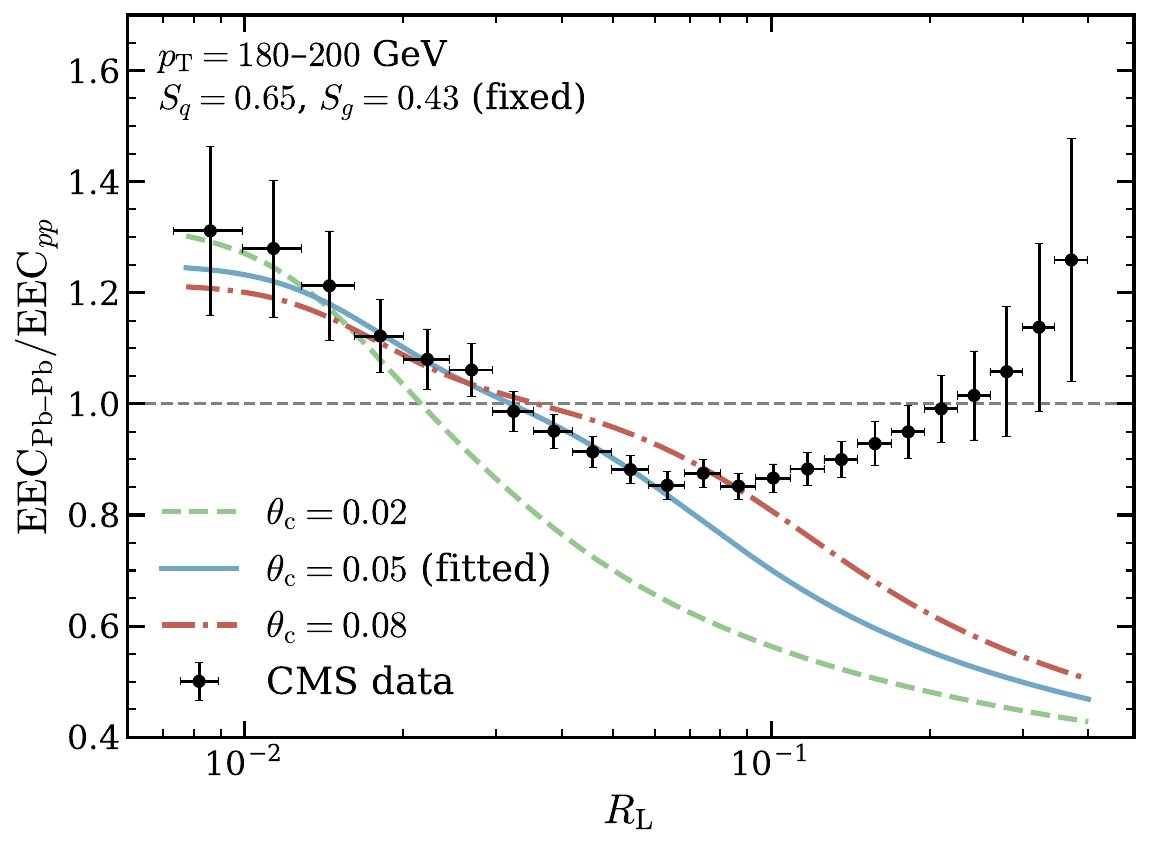}
    \caption{Ratio of the Pb--Pb to $pp$ EEC at $p_{\rm{T}} = 180$--$200$~GeV, compared to CMS data~\cite{CMS:2025ydi}, for three values of the decoherence angle $\theta_{\rm{c}}$. The quark and gluon quenching factors are held fixed at the values extracted from the joint fit to ATLAS $R_{\rm{AA}}$~\cite{ATLAS:2018gwx} and CMS EEC, from which $\theta_{\rm{c}}  \approx 0.05$ was obtained; only $\theta_{\rm{c}}$ is varied.}
    \label{fig:eec_ratio_theta_c}
\end{figure}

The sensitivity of the fit to $\theta_{\rm c}$ is investigated. Figure~\ref{fig:eec_ratio_theta_c} shows the Pb--Pb to $pp$ EEC ratio in the $180$--$200$~GeV window computed at $\theta_{\rm c} = 0.02$, $0.05$ and $0.08$, with $\Qc_q$ and $\Qc_g$ held at the values the joint fit returns for that window; nothing is refitted, so the spread between the curves isolates the dependence on $\theta_{\rm c}$. Relative to the fitted value, $\theta_{\rm c} = 0.02$ costs $\Delta\chi^2 = 280$ and $\theta_{\rm c} = 0.08$ costs $21$. Both are therefore excluded by the data, which is consistent with the small uncertainties on $\theta_{\rm c}$ in Fig.~\ref{fig:thetaC}, about $10\%$ in the highest $p_{\rm T}$ window and $20\%$ in the lowest.

This sensitivity to $\theta_{\rm c}$ stems from the shape of the interpolation between the two limits. The dipole quenching factor derived in App.~\ref{app:dipole} approaches its decoherent limit as a power law, $(\theta_{\rm c}/\theta)^{2/3}$, rather than exponentially, so an angle as small as
$\theta_{\rm c} \simeq 0.05$ still shapes the EEC at angles several times larger.

\subsection{Imposing Casimir scaling}
\label{app:casimirfit}

In the main text, Casimir scaling is a test: $\Qc_q$ and $\Qc_g$ are fitted independently, and the
relation $\Qc_g = \Qc_q^{9/4}$ derived in App.~\ref{app:Casimir} is checked afterwards. Here it is imposed instead. This leaves one quenching parameter per window rather than two, with $\theta_{\rm c}$ still free in the resolved case. Nothing else about the model, the data or the fit is changed.

\begin{figure}[H]
    \centering
    \includegraphics[width=0.4\textwidth]{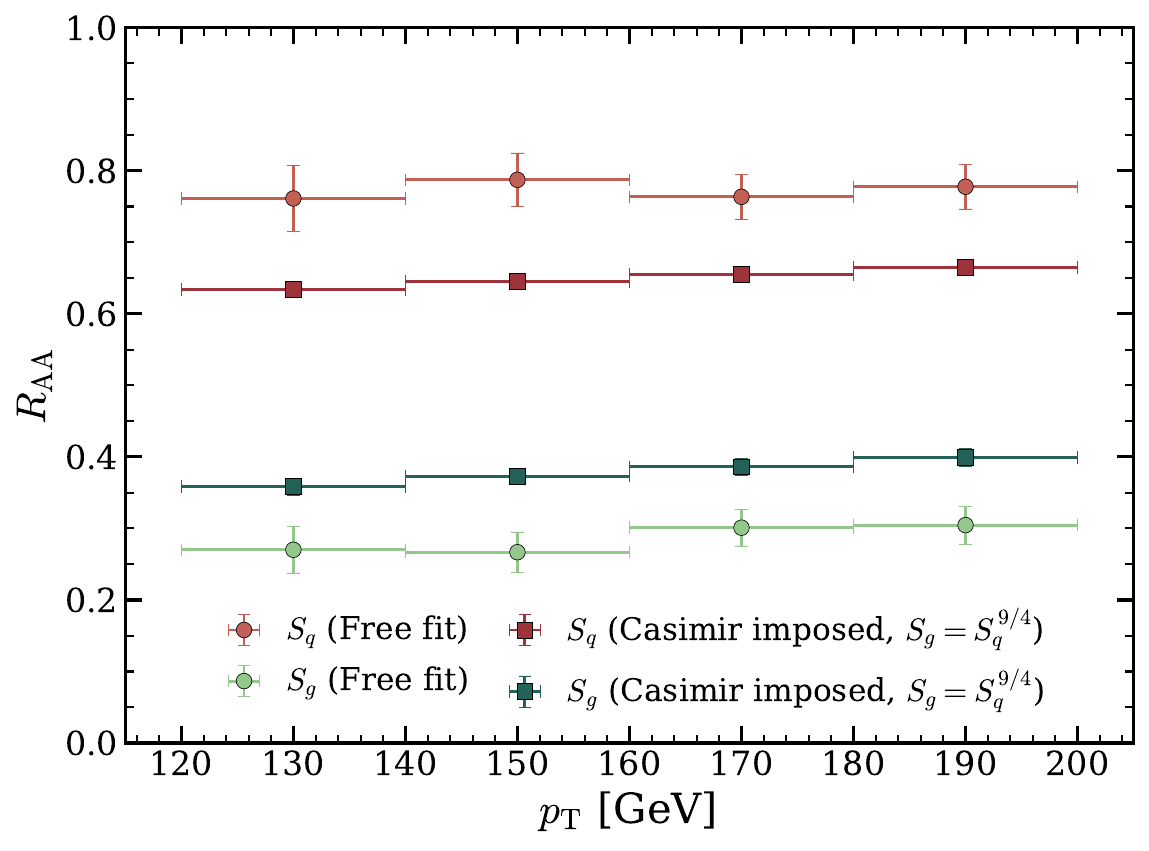}
    \caption{Quark and gluon quenching factors per $p_{\rm T}$ window from the fit in the coherent limit, extracted twice: once with $S_q$ and $S_g$ free and independent (circles), and once with the Casimir relation $S_g \equiv S_q^{9/4}$ imposed (squares). Here the two versions separate: left free, the coherent fit drives the gluon factor down to $S_g = 0.27$--$0.30$, far below the Casimir value $S_q^{9/4} = 0.54$--$0.58$ that the constrained fit is
    forced onto. Imposing the relation costs $\Delta\chi^2 = 53.6$ for four fewer parameters, a $6.5\sigma$
    preference for the unconstrained split, and the constrained fit is left describing the EEC poorly. This is not the case once color-decoherence corrections are carried in the energy-loss expression, where the
    unconstrained and constrained fits coincide~(Fig.~\ref{fig:raa_casimir_vs_free_decoherent}).}
    \label{fig:raa_casimir_vs_free_coherent}
\end{figure}

\begin{figure}[H]
    \centering
    \includegraphics[width=0.4\textwidth]{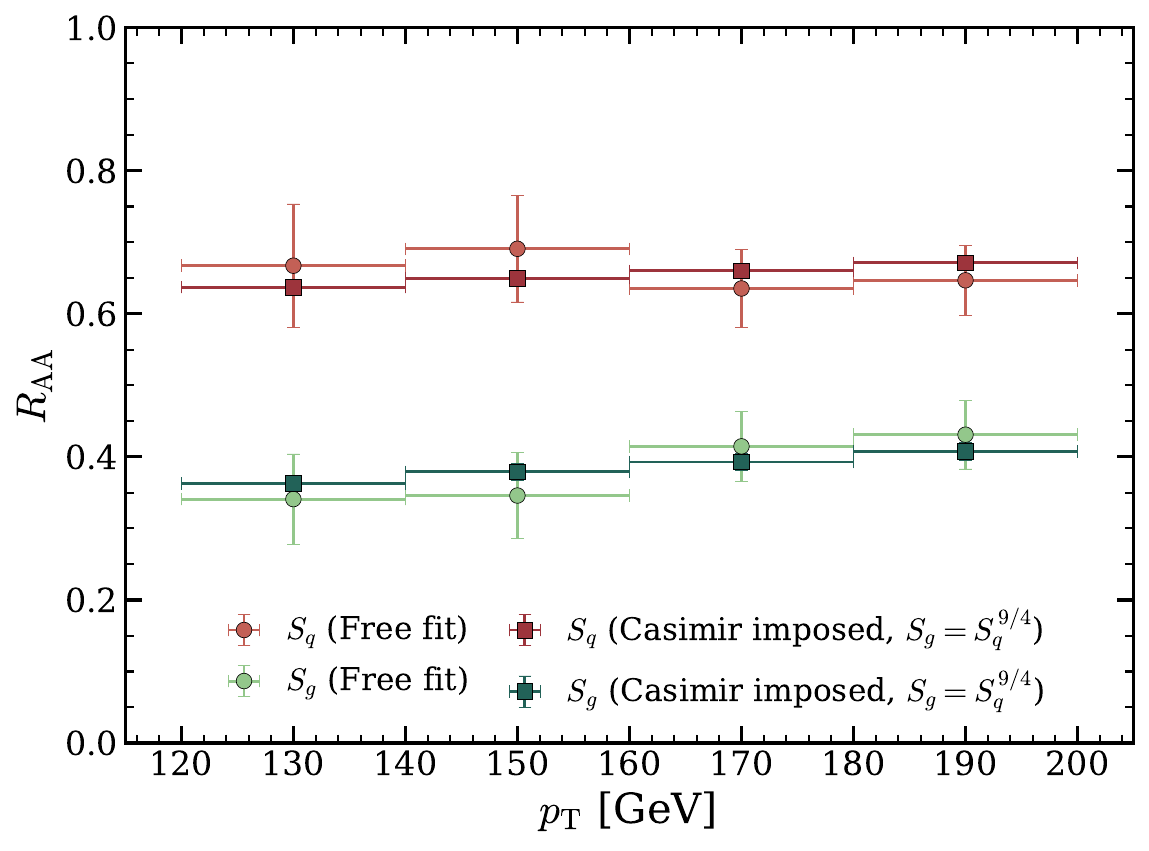}
    \caption{Quark and gluon quenching factors per $p_{\rm{T}}$ window from the fit including color-decoherence corrections, extracted twice: once with $S_q$ and $S_g$ free and independent~(circles), and once with the Casimir relation $S_g \equiv S_q^{9/4}$ imposed (squares). The two agree window by window: with the decoherence correction carried in the energy-loss expression, the unconstrained fit lands on the Casimir line by itself. This is not the case in the coherent approximation as shown in Fig.~\ref{fig:raa_casimir_vs_free_coherent}.}
    \label{fig:raa_casimir_vs_free_decoherent}
\end{figure}

In the coherent limit, the constraint costs $\Delta\chi^2 = 53.6$ for four fewer parameters. That is a $6.5\sigma$ preference for leaving $\Qc_g$ free, and the constrained and free fits are far apart in every window, as shown in Fig.~\ref{fig:raa_casimir_vs_free_coherent}. Once color-decoherence corrections are included, the
same constraint costs only $\Delta\chi^2 = 0.90$. The two fits then overlap window by window as depicted in Fig.~\ref{fig:raa_casimir_vs_free_decoherent}, and the data cannot separate them. This is the conclusion of the main text, reached by refitting under the constraint rather than by comparing two extracted numbers. It is the more conservative statement, since the rest of the fit is free to react.

\begin{figure}[H]
    \centering
    \includegraphics[width=0.4\textwidth]{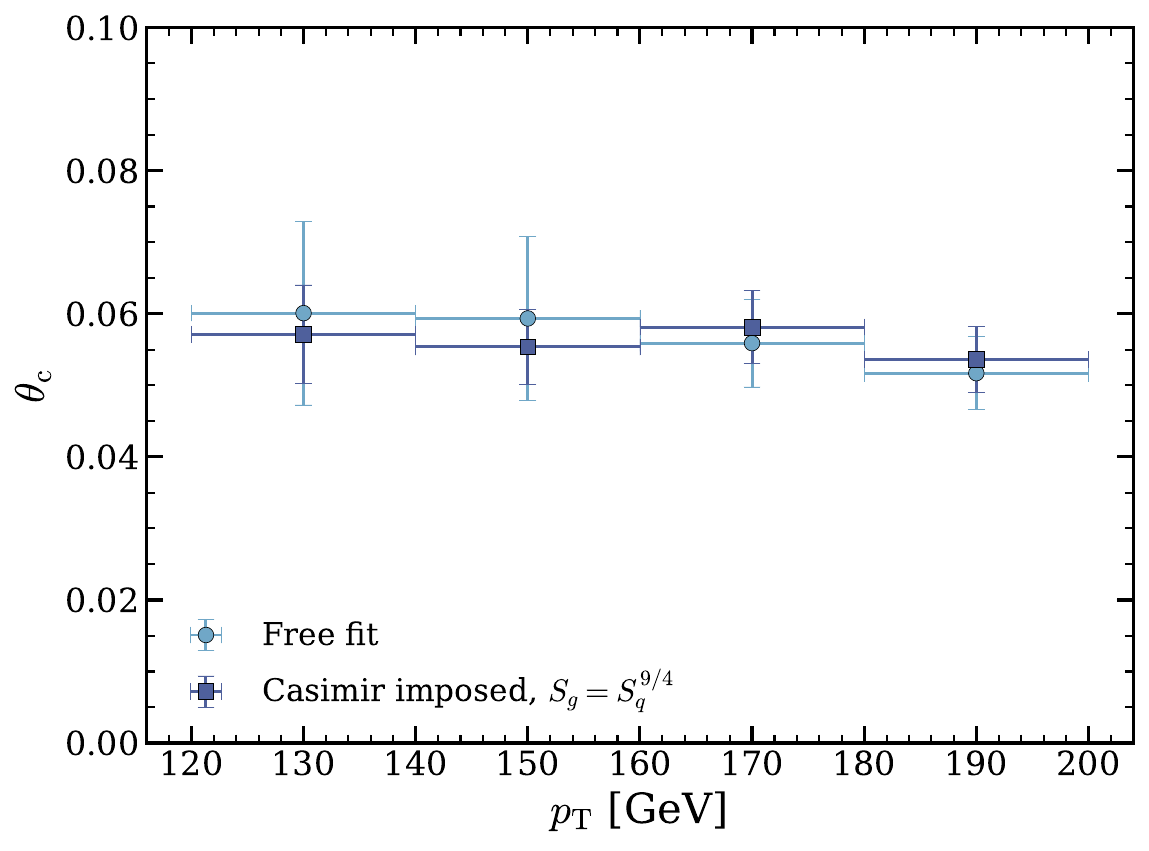}
    \caption{Decoherence angle $\theta_{\rm{c}}$ per $p_{\rm{T}}$ window, from the same two fits as in Fig.~\ref{fig:raa_casimir_vs_free_decoherent}. The angles agree within
    $1\sigma$ in every window and both stay flat at $\theta_{\rm{c}} \simeq 0.06$.}
    \label{fig:theta_c_casimir_vs_free}
\end{figure}

With color-decoherence corrections taken into account, $\theta_{\rm c}$ is left free to absorb the constraint. However, Figure~\ref{fig:theta_c_casimir_vs_free} shows that the constrained and free angles agree within $1\sigma$ in every window, with roughly half the error, as expected with one parameter fewer.

\subsection{Dependence on the upper angular limit of the fit}
\label{app:rlmax}

The dependence of the fit on its upper angular limit~$R_{\rm L}^{\max}$ is investigated. This is the only tunable element of the medium step. It is set to $0.07$ \emph{a priori}, to keep the fit inside the region where hard collinear modes dominate. The choice is corroborated \emph{a posteriori} by the fitted decoherence angle, which comes out just below it.

\begin{figure}[H]
    \centering
    \includegraphics[width=0.4\textwidth]{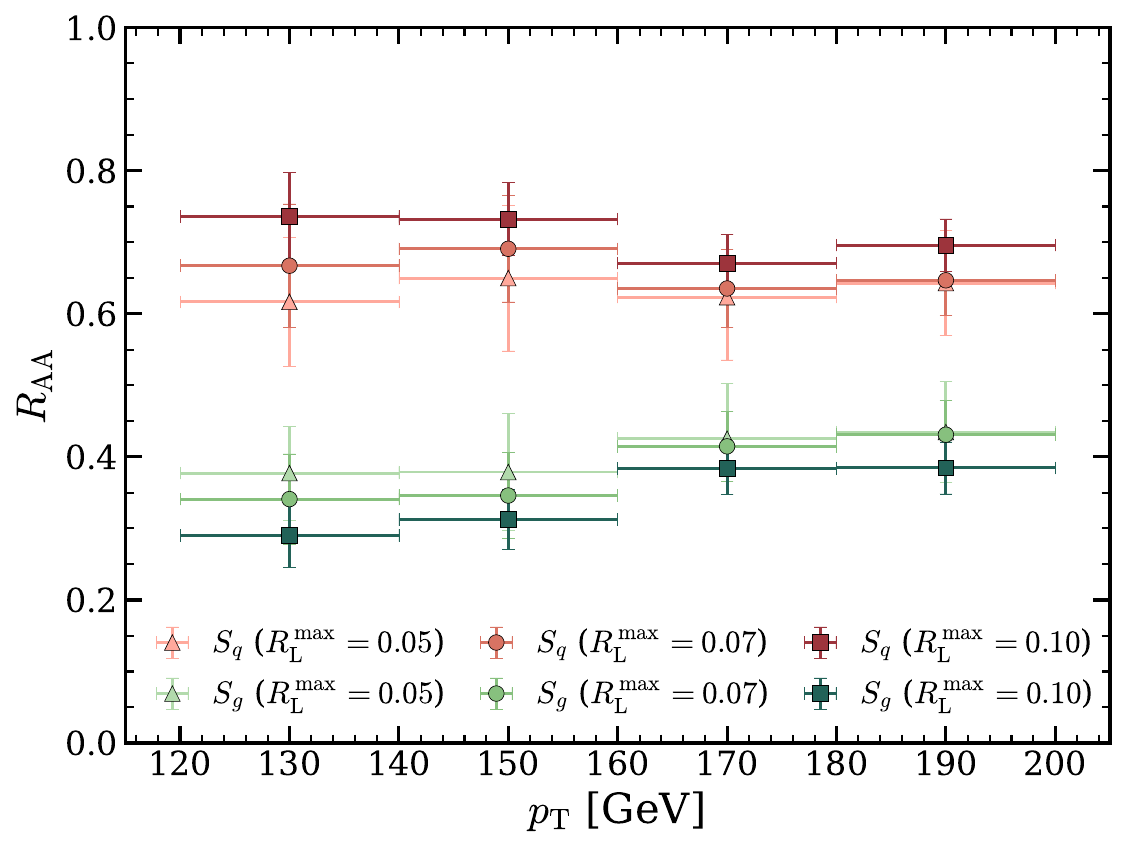}
    \caption{Quark and gluon quenching factors per $p_{\rm T}$ window from the joint fit to ATLAS $R_{\rm{AA}}$~\cite{ATLAS:2018gwx} and CMS EEC~\cite{CMS:2025ydi}, repeated with the Pb--Pb fit bounded by three different upper angular limits: $R_{\rm{L}}^{\max} = 0.05$ (triangles), $0.07$ (circles, the value used throughout) and $0.10$ (squares). The $pp$ reference fit and the flavor fractions are identical in the three fits, so the spread between the marker sets measures the sensitivity to the angular window of the Pb--Pb fit alone. The flavor hierarchy $S_g < S_q$ survives at every limit; opening the window pushes $S_q$ up and $S_g$ down, as larger angles bring in soft medium-induced radiation and medium response that the calculation does not describe and the fit absorbs into the gluon factor.}
    \label{fig:quenching_rlmax}
\end{figure}

\begin{figure}[H]
    \centering
    \includegraphics[width=0.4\textwidth]{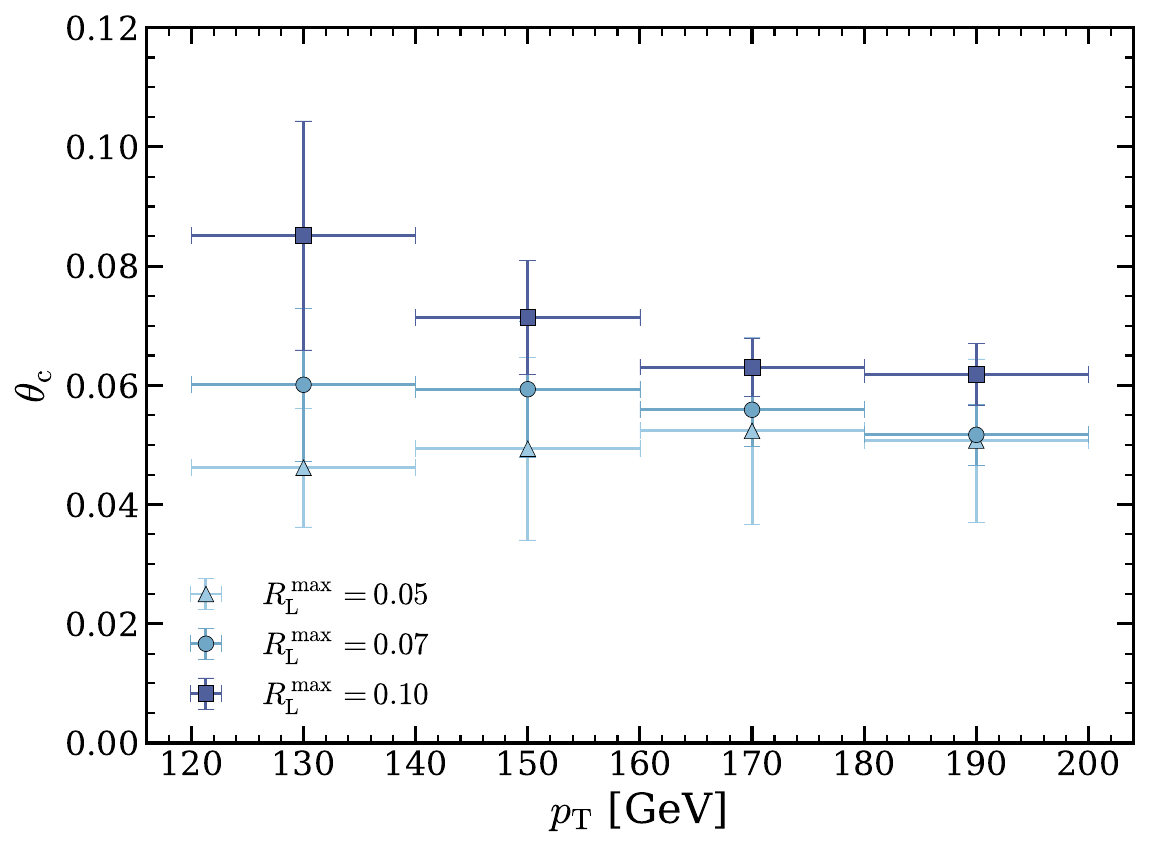}
    \caption{Decoherence angle $\theta_{\rm{c}}$ per $p_{\rm{T}}$ window from the same three fits as in Fig.~\ref{fig:quenching_rlmax}, that is with the Pb--Pb fit bounded at $R_{\rm{L}}^{\max} = 0.05$ (triangles), $0.07$ (circles) and $0.10$ (squares). The central value tracks the upper limit, never settling far below $R_{\rm L}^{\rm{max}}$, which is the expected behavior since the last fitted points carry the
    transition. That drift is small against the uncertainties, however: the three determinations agree within $1.8\sigma$ in the worst window and well within $1\sigma$ in most, and in the two highest-$p_{\rm T}$ windows, where $\theta_{\rm{c}}$ is measured most precisely, the $0.05$ and $0.07$ fits are indistinguishable. The extracted angle is therefore not strongly tied to the choice of $R_{\rm{L}}^{\max}$. The fit quality does
    degrade monotonically as the window opens, $\chi^2/$dof $= 0.12$, $0.13$, $0.22$, and
    $R_{\rm{L}}^{\max} = 0.07$ is the compromise adopted in this work.}
    \label{fig:theta_c_rlmax}
\end{figure}

The medium step is repeated at $R_{\rm L}^{\max} = 0.05$ and $0.10$, a factor of two apart. The vacuum
reference and the flavor fractions are left untouched, so the spread between the three sets measures the angular window alone. Opening the window from $0.05$ to $0.07$ to $0.10$ takes $\chi^2/$dof from $0.12$ to $0.13$ to $0.22$. The quenching factors and the decoherence angle shift slightly with it, as shown in Figs.~\ref{fig:quenching_rlmax} and~\ref{fig:theta_c_rlmax}.

Two things follow. First, the result of the Letter is stable under a change of the upper angular limit: the flavor hierarchy holds at every limit, and the resolved fit stays compatible with Casimir scaling. Second, the fit quality degrades monotonically as the window opens. This is expected of a calculation pushed past its region of validity, not of a mistuned parameter. We therefore report the spread across these limits as the systematic uncertainty on $\theta_{\rm c}$, rather than the per-window fit errors,
\begin{equation*}
    \theta_{\rm c} = 0.057 \pm 0.005~({\rm fit})\,^{+0.014}_{-0.008}~({\rm syst})\,,
\end{equation*}
the window-averaged angle moving $0.049 \to 0.057 \to 0.071$ over the three limits. The corresponding spread on $\Qc_q$ and $\Qc_g$ is about $\pm 0.05$.

\subsection{Compatibility of measured distributions with Casimir scaling}\label{app:Casimir}

Quarks and gluons are expected to lose energy differently as a result of their different color charges. The expected energy loss difference between quarks and gluons can be expressed in terms of the Casimir color factors where $C_{F} = 4/3$ for quarks and $C_{A} = 3$ for gluons. In this appendix, we discuss the compatibility of the extracted $R_{\rm AA}$ distributions with the expectations of Casimir scaling. If the energy loss distributions follow Casimir scaling, we would expect the energy loss to be given as 
\begin{equation} \label{eq:QG}
    C_{F}\Delta E_{g} = C_{A}\Delta E_{q}
\end{equation}

\noindent where $\Delta E_{g}$ and $\Delta E_{q}$ are the energy loss distributions for gluons and quarks, respectively. However, we measure $R_{\rm AA}$, which is distinctly different from the $\Delta E$. Therefore, we need to map the conditions imposed on the $\Delta E$ to the $R_{\rm AA}$. To do this, we begin with the expression for the $R_{\rm AA}$ from Eq.~(8) of Ref.~\cite{Spousta:2015fca}, while using the variable $L(p_{\rm T})$ instead of $S(p_{\rm T})$ to avoid confusion between this variable and the $S_{q,g}$ in the main text of the paper. Making that substitution gives
\begin{equation}\label{eq:StartRaa}
    R_{\rm AA} (p_{\rm T}) = \left(\frac{1}{1 + L(p_{\rm T})/p_{\rm T}}\right)^{n} \left(1 + \frac{\rmd L}{\rmd p_{\rm T}}\right),
\end{equation}
where $n$ is the power-law index of the jet spectrum. For convenience, let us define the fractional shift such that $\sigma \equiv \frac{L(p_{\rm T})}{p_{\rm T}}$. Additionally, from Eq.~(18) of Ref.~\cite{Spousta:2015fca} we see that 
\begin{equation}
    L = \ell' \left(\frac{p_{\rm T}}{p_{\rm T, 0}}\right)^{\alpha},
\end{equation}
where $\alpha$ is the exponent in the energy-loss scaling. To write the Jacobian term, $\frac{dL}{dp_{\rm T}}$, in terms of $\sigma$, we get
\begin{equation}
    \frac{\rmd L}{\rmd p_{\rm T}} = \frac{\rmd}{\rmd p_{\rm T}}\left[\ell' \left(\frac{p_{\rm T}}{p_{\rm T, 0}}\right)^{\alpha}\right] = \ell' \frac{\alpha}{p_{\rm T}}\left(\frac{p_{\rm T}}{p_{\rm T, 0}}\right)^{\alpha}  = \alpha \sigma.
\end{equation}

\noindent Then Eq.~\eqref{eq:StartRaa} becomes
\begin{equation}
       R_{\rm AA} (p_{\rm T}) = (1+\sigma)^{-n} (1 + \alpha\sigma).
\end{equation}
\noindent Taking the natural logarithm of both sides and taking the limit where the $\sigma \ll  1$, we get 
\begin{equation}\label{eq:lnRaa}
    \ln( R_{\rm AA} (p_{\rm T})) = -(n-\alpha)\sigma + \mathcal{O}(\sigma^{2}).
\end{equation}

\noindent Now, to enforce Casimir scaling at a fixed $p_{\rm T}$, we get 
\begin{equation}
    \sigma_{g} = \frac{C_{A}}{C_{F}}\sigma_{q}. 
\end{equation}

\noindent Substituting in Eq.~\eqref{eq:lnRaa}, we get
\begin{equation}\label{eq:Casimir2}
    \frac{\ln( R^{g}_{\rm AA} (p_{\rm T}))}{-(n-\alpha)} = \frac{C_{ A}}{C_{ F}} \frac{\ln( R^{q}_{\rm AA} (p_{\rm T}))}{-(n-\alpha)}. 
\end{equation}

\noindent Here, for simplicity, we assume that $n$ is approximately equivalent for quarks and gluons~\footnote{Note that this is not strictly the case: quarks have a slightly steeper spectrum than gluons. However, for the purposes of this consistency check they can be treated as approximately equivalent.}. Then Eq.~\eqref{eq:Casimir2} becomes 
\begin{equation}\label{eq:Casimir3}
   \ln( R^{g}_{\rm AA} (p_{\rm T})) = \frac{C_{ A}}{C_{F}} \ln( R^{q}_{\rm AA} (p_{\rm T}))
\end{equation}
or in other words
\noindent \begin{equation}\label{eq:Casimir4}
   R^{g}_{\rm AA} (p_{\rm T}) = (R^{q}_{\rm AA} (p_{\rm T}))^{\frac{C_{ A}}{C_{ F}} } = (R^{q}_{\rm AA} (p_{\rm T}))^{9/4}
\end{equation}
\noindent which is the final result for the relationship between the $R_{\rm AA}$ for quark- and gluon-initiated jets in the event that Casimir scaling holds. The values obtained from this relation in the fully coherent case are given in App.~\ref{app:fit-res}, and the compatibility of the resolved case is discussed in the main text.

\bibliographystyle{apsrev4-1}

\bibliography{letter.bib}

\end{document}